\documentclass[a4paper,fleqn, 11pt]{cas-sc}

\usepackage{upgreek}
\usepackage{subcaption}
\usepackage{soul}
\usepackage{amsmath,bm}
\usepackage{flafter} 
\usepackage[numbers]{natbib}
\usepackage{float}
\usepackage[normalem]{ulem}
\usepackage{placeins}
\usepackage{tcolorbox}

\DeclareMathOperator*{\argmax}{arg\,max}%
\DeclareMathOperator*{\argmin}{arg\,min}%
\newcommand{\g}{\boldsymbol{g}}%
\newcommand{\A}{\boldsymbol{a}}%

\newcommand{\hz}{\hat{z}}%
\newcommand{\hb}{\hat{b}}%
\newcommand{\e}{\varepsilon}%

\renewcommand{\vec}[1]{\boldsymbol{#1}}

\newcommand \dd[1]  { \,\textrm d{#1}}   

\def\tsc#1{\csdef{#1}{\textsc{\lowercase{#1}}\xspace}}
\tsc{WGM}
\tsc{QE}
\tsc{EP}
\tsc{PMS}
\tsc{BEC}
\tsc{DE}

\begin{document}
\let\WriteBookmarks\relax
\def\floatpagepagefraction{1}
\def\textpagefraction{.001}
\shorttitle{Shell theory of the cell cortex}
\shortauthors{Hudson Borja da Rocha et~al.}

\title [mode = title]{A viscous active shell theory of the cell cortex}                      


\author[1]{{\color{black}Hudson Borja da Rocha}}[orcid = 0000-0001-9615-2022
]

\address[1]{Center for Interdisciplinary Research in Biology (CIRB), Collège de France, CNRS, INSERM, Université PSL, Paris, France}

\author[2]{{\color{black}Jeremy Bleyer}}[orcid = 0000-0001-8212-9921
]

\address[2]{Laboratoire Navier, Ecole des Ponts ParisTech, Université Gustave Eiffel, CNRS, 
6-8 av Blaise Pascal, Cité Descartes, 77455 Champs-sur-Marne, FRANCE }

\author[1]{{\color{black}Hervé Turlier}}[orcid = 0000-0002-9332-9125
]
\cormark[1]

\cortext[cor1]{Corresponding author. \\
 \textit{E-mail addresses:} \href{mailto:hudson.borja-da-rocha@polytechnique.edu}{hudson.borja-da-rocha@polytechnique.edu} (H. Borja da Rocha),  \href{mailto:jeremy.bleyer@enpc.fr}{jeremy.bleyer@enpc.fr} (J. Bleyer), \href{mailto:herve.turlier@college-de-france.fr}{herve.turlier@college-de-france.fr} (H. Turlier)}
%

\begin{abstract}
The cell cortex is a thin layer beneath the plasma membrane that gives animal cells mechanical resistance and drives most of their shape changes, from migration, division to multicellular morphogenesis. It is mainly composed of actin filaments, actin binding proteins, and myosin molecular motors. Constantly stirred by myosin motors and under fast renewal, this material  may be well described by viscous and contractile active-gel theories. Here, we assume that the cortex is a thin viscous shell with non-negligible curvature and use asymptotic expansions to find the leading-order equations describing its shape dynamics, starting from constitutive equations for an incompressible viscous active gel. Reducing the three-dimensional equations leads to a Koiter-like shell theory, where both resistance to stretching and bending rates are present. Constitutive equations are completed by a kinematical equation describing the evolution of the cortex thickness with turnover. We show that tension and moment resultants depend not only on the shell deformation rate and motor activity but also on the active turnover of the material, which may also exert either contractile or extensile stress. Using the finite-element method, we implement our theory numerically to study two biological examples of drastic cell shape changes: osmotic shocks and cell division. Our work provides a numerical implementation of thin active viscous layers and a generic theoretical framework to develop shell theories for slender active biological structures.
\end{abstract}

%


\maketitle

\section{Introduction}
Biological cells can exhibit a large diversity of shapes, which are intimately linked to the multiple key functions they perform: division, polarisation, apoptosis (i.e., programmed death), or migration. Dynamic changes in the shape of animal cells, also called \textit{morphogenesis}, is mainly controlled by the cell cortex, a dense meshwork of filaments underlying the plasma membrane, capable of generating internal active tensions and torques \cite{clark2014stresses}. 

The cell cortex is a thin biological structure of few hundreds of nanometers in thickness that extends over several tens of micrometers, the typical cell size \cite{clark2013monitoring}. It is composed of semi-flexible actin polymers, actin-binding proteins, and myosin molecular motors, which are the substrates  for various phosphorylation events keeping this material out of equilibrium \cite{kelkar2020mechanics}. 
By continuously hydrolyzing ATP (adenosine triphosphate) into ADP (adenosine diphosphate) and inorganic phosphate, actomyosin networks have two remarkable properties, which are relatively unusual in classical mechanics. 
First, they are under permanent renewal, principally through the assembly and disassembly of filaments, which happens within seconds; this fast material turnover releases elastic stresses accumulated in the network, making this material viscous on timescales longer than a few dozens of seconds \cite{Khalilgharibi:2019jt}. Second, they are the subject of internal active loads created by molecular motors, which exert contractile traction forces on neighboring filaments; this creates a source of active stress that is continuously dissipated and renewed, leading overall to an effective surface tension at the cell level \cite{tinevez2009role,mayer2010anisotropies, Turlier_BioPhys_2014, Maitre2016}; it can also lead to structural rearrangements of actin filaments within the plane of the cell's surface, which has motivated the recent emergence of active-gel theories \cite{Prost_Nature_2015}. Relying on mass and momentum balance and exploiting the fundamental symmetries of the system, active-gel theories extend hydrodynamic theories for nematic or polar liquid crystals \cite{Prost_1993} to account for the activity of motors and, to a certain extent, for the renewal of the material \cite{Kruse_EPJE_2005, Prost_Nature_2015}. They have provided a convenient generic theoretical framework to describe the large-scale mechanical properties of active materials through hydrodynamic variables. The cortex mechanics is consecutively described in general by two sources of stress: an active stress, which is controlled spatiotemporally by the cell metabolism and depends on the local filament organization, and a passive stress, which results from the viscoelastic deformation of the network in response to external loads and internal active stress. This theoretical framework has proven its ability to capture the essential physics characterizing the morphological changes observed experimentally in single cells \cite{tinevez2009role, mayer2010anisotropies, Turlier_BioPhys_2014, Salbreux_PRL_2009, Hawkins:2011kl, Bun:2014ei, CallanJones:2016fe, malik2019scaling}.

Because of the slender character of the cell cortex (around 0.2--2$\mu m$ in thickness \cite{hiramoto1957thickness,clark2013monitoring} for a typical cell radius of 10--100$\mu m$), it is reasonable to adopt thin shell theories to model its mechanics at the cell level \cite{Berthoumieux:2014eo, Yin:2021gw}. Shells are considered thin if their thickness is much smaller - typically by one order of magnitude or more - than other characteristic dimensions, such as the curvature radius. The main goal of thin shell theories is to reduce the three-dimensional equations into an equivalent two-dimensional model, which is generally defined onto the midsurface (i.e. the surface that splits the shell into two layers of equal thickness at any point), although different choices of reference surfaces are possible \cite{petrov1984elastic,morris2019active}.  
To perform this dimensional reduction, two generic approaches are usually undertaken: one may calculate the limit, as the slenderness parameter goes to zero, of the the nonlinear 3D model using asymptotic methods; alternatively one may restrict, from the beginning, the range of admissible deformations and stresses used in the nonlinear 3D shell model, by means of - generally \textit{ad hoc} - assumptions, for instance the Kirchhoff-love hypothesis \cite{love1888}.
Even though the passive response of the cortex is essentially viscous at the timescale of interest, we expect here our viscous active shell model to be closely related elastic plates and shell theories, which have been extensively studied in solid mechanics.  This is due to the Stokes-Rayleigh analogy, which states that the equations of slow viscous flow and linear elasticity are interchangeable in the incompressible limit if velocities replace displacements and strain rates replace strains \cite{Stokes_1845, Rayleigh_1945,batty2012discrete,bhattacharyastokes}. For a more detailed introduction on elastic shell theories, we refer to classic textbooks \cite{niordson2012shell, GreenZerna, Bischoff_2017, Chapelle_FEM_2011}. For viscous materials, only a limited number of thin sheet models have been previously derived. A few membrane models (in which bending resistance is neglected) have been developed in arbitrary shapes \cite{Fliert_JFM_1995,Howell_EPAM_1996} but models accounting for bending resistance were mostly derived for specific geometries \cite{buckmaster1975buckling, Ribe_JFM_2001, Perdigou2016}, with one exception in general curvilinear coordinates \cite{Ribe_JFM_2002}. Relying on an asymptotic approach, this last study constitutes the starting point of our active shell theory for the cell cortex.

\begin{figure}[ht]
\centering
\includegraphics[width=\textwidth]{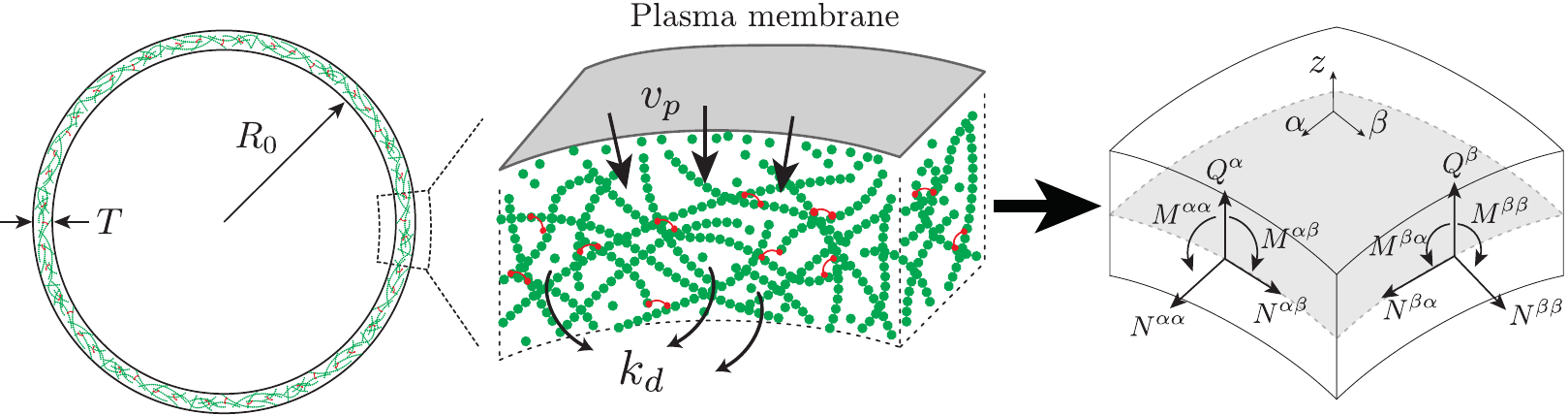}
\caption{Schematic representation of the cell cortex, composed of actin filaments and actin-binding proteins, in particular myosin molecular motors. The cortex forms a thin layer of thickness $T\ll R_0$ beneath the plasma membrane, where $R_0$ is the typical curvature radius of the cell. It is under permanent turnover, with a polymerization flux $v_p$ from the plasma membrane surface and a depolymerization rate $k_d$ in the bulk of the layer. The dimensional reduction into a thin theory theory consists in finding the two-dimensional approximate in-plane and transverse shear stress $N^{\alpha\beta}$ and $Q^{\alpha}$, and torque $M^{\alpha\beta}$ resultants, defined on the middle-surface of the shell.}\label{fig:cortex_full}
\end{figure}

Most of the recent works on dynamic cell shape changes have considered the mechanics of the cortex, modeled as an active and viscous surface equipped with tensions only \cite{yeung1989cortical, Turlier_BioPhys_2014, Mietke_PNAS_2019, Arroyo_JFM_2019}. 
Other works have focused on the lipid bilayer (or plasma membrane) mechanics (onto which the cortex is attached through specialized linker proteins), which is generally described as a 2-dimensional fluid layer equipped with an elastic bending rigidity \cite{schneider1984thermal,arroyo2009relaxation,rangamani2013interaction,fournier2015hydrodynamics,sauer2017stabilized,Sahu_PRE_2017}. Lipid bilayers may indeed be regarded as a stack of two lipid monolayers of nanometric thickness that do not flow transversely, which justifies why viscous dissipation is inherently 2-dimensional in that case. Here, we argue that the lipid bilayer contribution to cell mechanics is generally negligible, and we focus on the cortex. The cortex present several features that justify the need to derive a proper thin viscous shell theory starting from 3-dimensional constitutive equations. First, the cortex is a bulk continuous material, with a thickness that may vary along the cell's surface, which is in contrast to the classical assumption of constant thickness in thin elastic shell theories \cite{bischoff2004models} or lipid bilayers. Second, the turnover of cortical material is not homogeneous along the transverse sheet direction: polymerization happens mainly at the lipid bilayer surface, through membrane-attached actin nucleator proteins called formins \cite{van2015plasma}, while depolymerization acts in bulk, idealistically at each filament ends \cite{fritzsche2013analysis}. As commonly described, we assume that this mode of turnover breaks the symmetry of the shell with respect to its midsurface by creating an inward flux of actomyosin toward the cell interior, as depicted schematically on Fig.~\ref{fig:cortex_full}. This flux can potentially generate tangential viscous flows of cortical material, and therefore and additional active source of stress, that has been neglected so far. Third, pure membrane models of the cell surface cannot handle compressive forces nor active torque generation, that may lead to surface buckling. Finally, if active-gel theories provide relevant bulk constitutive equations for cortical material, the mathematical form of equivalent two-dimensional stress and torque resultants after dimensional reduction is hard to guess readily. 

This paper proposes therefore a general thin active viscous shell theory that aims to predict how the shape of a cell responds to arbitrary loads. The load can be internal, such as gradients of active stress, or external, originating from the mechanical interaction of the cell with its environment. Extending to active-gel materials a previous covariant description of thin viscous shells \cite{Ribe_JFM_2002}, we present a systematic asymptotic expansion of governing equations in powers of a slenderness parameter $\varepsilon \equiv T_0/L$, where $T_0$ and $L$ are respectively the characteristic thickness and a length scale over which the flows varies across the sheet (for a cell, $L$ is typically its radius $R_0$, and the thickness is fixed by turnover $T_0 \sim \frac{v_p}{k_d}$, see Fig.~\ref{fig:cortex_full}). We start from three-dimensional active-gel constitutive laws and mass balance equations. The expansion is performed in two limits, stretching and bending-dominated, which are then combined to build an approximate two-dimensional model that is valid for the whole range of intermediate behaviors, following the generic method of matched asymptotic expansion \cite{Fraenkel:1969}. The two-dimensional model is complemented by a kinematic equation describing the evolution of the thickness. Aiming at facilitating the spread, modification, refinement of our model and to highlight its features, we furthermore devise a numerical scheme \cite{githubFENICS} (available on \href{https://github.com/VirtualEmbryo/active_viscous_shell}{https://github.com/VirtualEmbryo/active$\_$viscous$\_$shell}) to solve the 2-dimensional thin shell equations with the finite-element library FEniCS \cite{alnaes2015fenics}, and we illustrate numerically two drastic dynamic cell shape changes: cell division and osmotic shocks.

The paper is organized as follows. In section \ref{sec:diffgeom-kinematics}, we start by introducing our notations and provide a summary of differential geometry and kinematics for surfaces and shells. In section \ref{sec:governingequations}, we introduce the general governing equations: the linear momentum balance in three dimensions and resulting generic stress resultants and torque balance on a surface; the three-dimensional constitutive equation of the cell cortex, and the mass balance with turnover. Follows in section \ref{sec:thinsheetequations} the core of the theory, with the dimensional reduction of constitutive equations by asymptotic expansion, which we perform successively in the membrane and inextensional limits, based on preliminary scaling arguments. Then we introduce a variational formulation of our problem in section \ref{sec:variationalform}, and we derive the kinematic evolution equation for the thickness in section \ref{sec:thicknessevolution}. We summarize the main equations of the model in section \ref{sec:summaryModel} and detail the numerical scheme employed to solve 2-dimensional thin sheet equations in section \ref{sec:numericalscheme}. We illustrate numerical results for osmotic shocks and cell division in section \ref{sec:numericalillustration} and discuss in the light of future perspectives to extend our theoretical and numerical approaches in the Dicussion, section \ref{sec:conclusion}. The Appendix contains more detailed calculations and numerical values used for simulations.

\section{Differential geometry and kinematics of surfaces and shells}\label{sec:diffgeom-kinematics}

Throughout the text, we use the Einstein summation convention: when an index variable appears twice in an expression, it implies summation of that variable over all values of the index.\footnote{The index appears only twice, once as a superscript (upper index $x^i$), and once as subscript (lower index $x_i$).} Latin indices take the value 1, 2, 3, and are used in variables describing the ambient, three-dimensional, Euclidean space. Greek indices take the value 1, 2, and are used in variables in the embedded, two-dimensional, surface. Therefore,  $x_i y^i=x_1 y^1+x_2 y^2+x_3 y^3$ and $s_\alpha s ^\alpha = s_1s^1+s_2 s^2 $. Bold symbols represent vectors. For a more complete text about differential geometry of surfaces and tensor calculus, we refer the reader to \cite{grinfeld2013tensorcalculus}; and to \cite{simmonds2012brief} for a concise introduction to tensor analysis.

\subsection{Differential geometry of surfaces}

\begin{figure}[ht]
\centering
\includegraphics[scale=1]{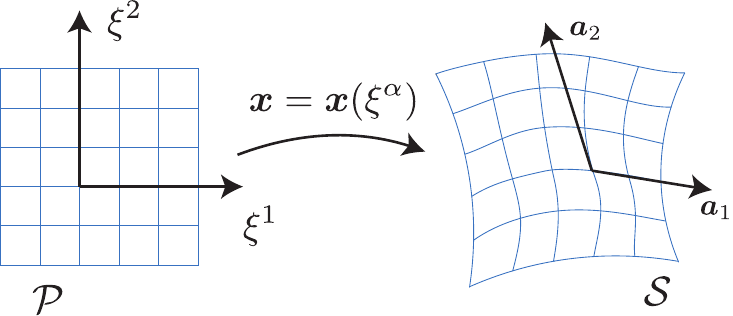}
\caption{Parametric description of a surface patch: each set of coordinates $(\xi_1,\xi_2)$ in the parametric domain \begin{math}\mathcal{P}\end{math} is mapped to a unique a material point $\bm{x}$ on the surface \begin{math}\mathcal{S}\end{math}} \label{fig:patch}
\end{figure}

Let $ \mathcal{S}$ be a surface embedded in $\mathbb{R}^3$ (e.g. the middle surface of the shell). We describe surfaces parametrically, that is, we choose a set of coordinate variables $(\xi^1, \xi^2)$ in a parametric domain $\mathcal{P}\in \mathbb{R}^2$ and let each of the ambient coordinates $x_1,x_2,x_3$ of a material point $\bm{x}\in \mathcal{S}$ depend on $\xi^\alpha$: $\bm{x}=\bm{x}(\xi^\alpha)$. This corresponds to a mapping of the point $(\xi^1,\xi^2)$ in the parametric domain $\mathcal{P}$ to a point on the material domain $\mathcal{S}$, see Fig.~\ref{fig:patch}. This mapping is supposed to be an infinitely differentiable homeomorphism. $\xi^\alpha$ are curvilinear coordinates on the surface $\mathcal{S}$ and are non-orthogonal in general. We must, therefore, distinguish between covariant and contravariant components of tensors. At every point in the surface of the patch, the parametrization $\xi^\alpha$ defines in-plane (covariant) tangent vectors
\begin{equation}
\vec{a}_\alpha=\bm{x}_{,\alpha}=\frac{\partial \bm{x}(\xi^\alpha)}{\partial \xi^\alpha},
\end{equation}
the comma denotes the partial derivative with respect to a variable. For tangential vectors to form a covariant vector basis $\vec{a}_{\alpha}$, we furthermore suppose them linearly independent at any point of the surface. The contravariant basis vectors $\vec{a}^{\alpha}$ are the reciprocals of the covariant basis, and are defined by the relation
$\vec{a}^{\alpha}\cdot \vec{a}_{\beta}=\delta^\alpha_\beta$, where $\delta^\alpha_\beta$ is the Kronecker delta.
Covariant and contravariant metric tensors of the surface are defined from basis vectors as $a_{\alpha\beta}=\vec{a}_{\alpha}\cdot\vec{a}_{\beta}$, and $ a^{\alpha\beta}=\vec{a}^{\alpha}\cdot\vec{a}^{\beta}$, respectively. These tensors allow one to generalize the notion of dot product on curved surfaces and may also be employed in lowering and raising index of tensors, for instance, $c^\alpha_\beta=a^{\lambda\alpha}c_{\beta\lambda}=a_{\beta\lambda}c^{\lambda\alpha}$. The determinant of the covariant metric tensor is $ a=\det(\A_{\alpha\beta})=a_{11}a_{22}-a_{12}a_{21}$. The unit normal of the surface $\mathcal{S}$ at any point $\bm{x}$ is defined by 
\begin{equation}
\vec{n} \equiv \frac{\vec{a}_1\times\vec{a}_2}{\left| \vec{a}_1\times\vec{a}_2 \right|}.
\end{equation}
While the normal is a unit vector, the same is not true for the tangent vectors, which generally are neither of unit length nor perpendicular to each other.

The second fundamental form of the surface is called the curvature tensor, defined as
$b_{\alpha\beta} \equiv -\A_\alpha\cdot \vec{n}_{,\beta}$, and is symmetric $b_{\alpha\beta}=b_{\beta\alpha}=-\A_\beta\cdot \vec{n}_{,\alpha}$. It quantifies the extent of local curvature of the surface in the embedding three-dimensional space.

In a curved manifold, the coordinate system is position dependent as well as the tensors living on this manifold. The covariant derivative accounts for the position dependence of the coordinate system, and extends the notion of derivatives to curved surfaces, producing tensors out of tensors. The resulting tensor is one covariant order greater than the original tensor. Covariant differentiation works differently for covariant and contravariant components and takes the following form,
\begin{subequations}\label{eq:CovariantDerivative}
\begin{align}
& v^\beta\vert_\alpha=v^\beta_{,\alpha}+\Gamma^\beta_{\alpha \gamma}v^\gamma, \\
&v_\beta\vert_\alpha=v_{\beta,\alpha}-\Gamma^\gamma_{\beta\alpha}v_\gamma.
\end{align}
\end{subequations}
where we introduced the Christoffel symbols, defined by $\Gamma^{\alpha}_{\lambda\mu} \equiv \A^{\alpha}\cdot \A_{\lambda,\mu}=\frac{1}{2}a^{\alpha\delta}(a_{\delta\lambda,\mu}+a_{\delta\mu,\lambda}-a_{\lambda\mu,\delta})$\footnote{Note that this definition of the Christoffel symbols suppose a torsion-free manifold (Levi-Civita connection)}. When acting on invariants\footnote{Objects that can be constructed without reference to any coordinate systems are called \textit{invariants} or \textit{geometric}. Scalar fields, vectors, and tensors are examples of geometric objects}, the covariant derivative is equivalent to the usual partial derivative, in particular $\vec{v}\vert_\alpha=\vec{v}_{,\alpha}$. Particular covariant derivatives are the ones of the basis vectors, which are known as Gauss and Weingarten's relations, given respectively by $\vec{a}_{\alpha\vert\beta} = b_{\alpha\beta}\vec{n}$, and $\vec{n}_{,\alpha} = - b^\lambda_\alpha \vec{a}_\lambda$.


    \subsection{Differential geometry of shells}
    
\begin{figure}[ht]
\centering
\includegraphics[scale=.8]{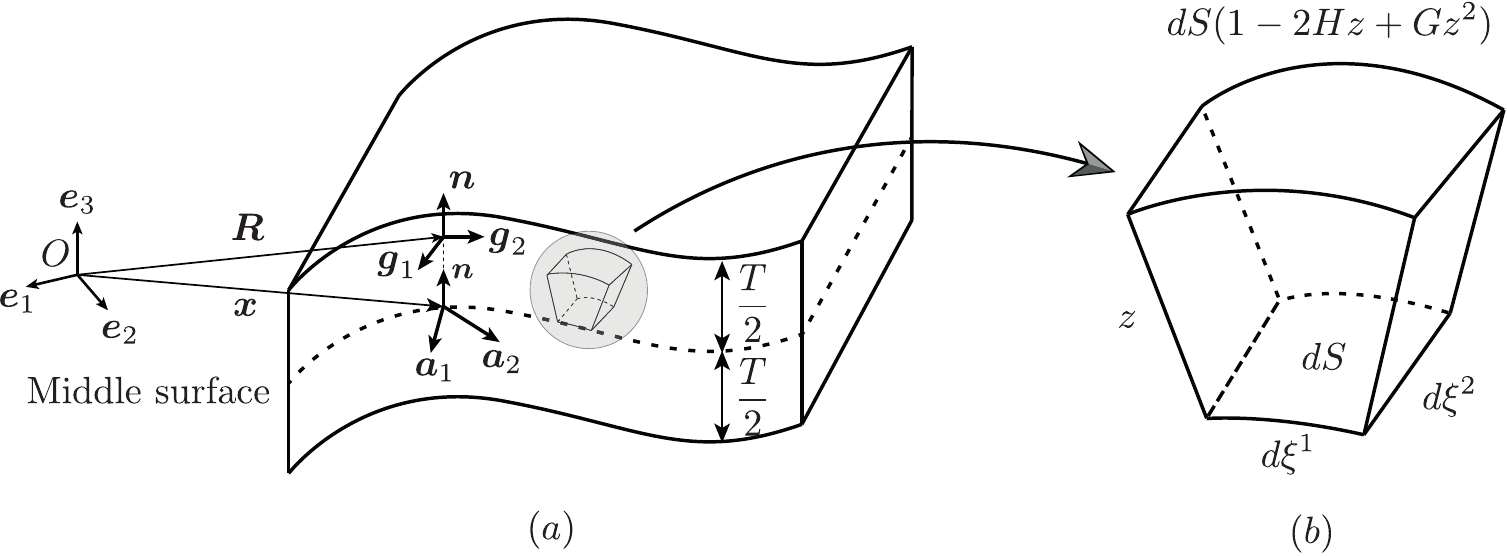}
\caption{(a) Parametrization of a shell of midsurface $\mathcal{S}$ and thickness $T$. The local basis vectors for arbitrary points in the shell are denoted by $\boldsymbol{g}_i$, while the basis vectors on the midsurface are $\boldsymbol{a}_\alpha$. The position of a material element in the shell is decomposed as $ \boldsymbol{R}(\xi^\alpha,z)=\bm{x}(\xi^\alpha)+z \vec{n}(\xi^\alpha)$, where $\bm{x}(\xi^\alpha,t)$ is the transverse projection onto the midsurface and $\boldsymbol{n}(\xi^\alpha)$ is the local normal to the midsurface. (b) Element of the shell: a patch of area $d\!S(0)$ at $z=0$ is multiplied by a factor $h= 1 - 2Hz + Gz^2$ at the transverse coordinate $z$ \label{fig:shell}}
\end{figure}

    Thin shells are usually described by a middle surface (or midsurface) $\mathcal{S}$ and a thickness $T$ \cite{GreenZerna,niordson2012shell}. The middle surface is the two-dimensional surface that is equidistant, at any of its point, to the upper and lower shell boundaries. The Euclidean space in which the surface is embedded is generally called the \textit{ambient} space. We suppose that the midsurface is everywhere continuous, infinitely differentiable and of arbitrary shape. In this text, we follow closely the notation of \cite{GreenZerna}, to which we refer the reader for a more in-depth discussion.

    The midsurface is described by a position vector $\bm{x}(\xi^\alpha,t)$ relative to an arbitrary origin through generalized coordinates $\xi^\alpha$. We introduce a third coordinate\footnote{Derivatives with respect to the normal coordinate will be denoted as $\partial \phi/\partial z\equiv\phi_{,3}$.}, $z \in [-T/2,T/2]$, measured along the unit normal $\vec{n}=\vec{a}_3(\xi^\alpha,t)$, defined at each point of the midsurface so that $z=0$ is the midsurface, and the thickness $T = T(\xi^\alpha)$ is a general function of $\xi^\alpha$, see Fig.~\ref{fig:shell}. Then, the position of an arbitrary material point of the shell $\boldsymbol{R}(\xi^\alpha,z,t)$ is defined by the equation
    \begin{equation}
    \label{eq:shellvectorposition}
        \boldsymbol{R}(\xi^\alpha,z,t)=\bm{x}(\xi^\alpha,t)+z \vec{n}(\xi^\alpha,t).
    \end{equation}

    The covariant basis vectors at an arbitrary point in the shell are defined as
    \begin{equation}
        \label{eq:galpha}
        \g_\alpha = \boldsymbol{R}_{,\alpha} = \A_{\alpha}-zb^{\lambda}_\alpha \A_\lambda=\mu^{\lambda}_\alpha \A_\lambda, \, \text{and }\,
        \g_3 = \vec{n}
    \end{equation}
    where $\mu^{\lambda}_\alpha=\delta^\lambda_\alpha-zb^\lambda_\alpha$ is a tensor that projects the ambient basis $\g_\alpha$ onto the midsurface basis $\A_\alpha$ (called \textit{shifter tensor} or \textit{shell shifter}). The determinant of the metric tensor is $g = \det g_{ij}$.
    The covariant metric tensor $g_{ij}=\vec{g}_i\vec{g}_j$ follows as 
    \begin{equation}
    \begin{split}
        g_{\alpha\beta}&=g_{\beta\alpha}=a_{\lambda\rho}\mu^\lambda_\alpha\mu^\rho_\beta\\
        g_{\alpha3}&=g_{3\alpha}=0\\
        g_{33}&=1
    \end{split}
    \end{equation}

    The contravariant basis vector at an arbitrary point in the shell is $\g^\alpha=h^{-1}(\mu^\rho_\rho\delta^\alpha_\lambda-\mu^\alpha_\lambda)\A^\lambda$, and $\g^3= \g_3 = \vec{n}$, where 
    \begin{equation}\label{eq:h}
        h=\sqrt{g/a} = \det(\mu^\beta_\alpha)=1-2H z+G z^2
    \end{equation}
    measures the deformation of a patch of area $dS$ at a distance $z$ from the midsurface because of shell curvature, and where 
\begin{equation}
    H=\frac{1}{2}b_{\alpha}^{\alpha} = \frac{1}{2}(b^1_1 + b^2_2), \qquad \text{and}\quad G=\det b^\beta_\alpha = b^1_1 b^2_2 - b^1_2 b^2_1,
\end{equation}
are the two main invariants of the curvature tensor, called the mean and Gaussian curvature, respectively. Finally, the contravariant metric tensor $g^{ij}=\vec{g}^i\vec{g}^j$ is such that $g^{ij} g_{j k} = \delta_{k}^i$.

\section{Governing equations}
\label{sec:governingequations}
The following development of the governing equations is standard and their derivation may be found in more detail in classical books on elastic shell theory~\citep{niordson2012shell, GreenZerna}. We adopt mostly the notations employed by \citep{GreenZerna}, to which we refer for more details.

    At the scale of cells, the Reynolds number is very small, and we are interested in deformations at low frequencies, such that inertial and convective terms in the Navier-Stokes equation are negligible \cite{happel2012low}. Therefore, all equations below will be written neglecting inertia. We assume that the distribution of stress throughout the sheet is characterized by the stress tensor $\vec{\tau}$, in equilibrium with bulk forces $\vec{f}$, described by the inertialess Cauchy's equation of motion
    \begin{equation}
        \text{div} \vec{\tau} + \rho \vec{f} = 0.
    \end{equation}
    The Voss-Weyl formula \cite{grinfeld2013tensorcalculus} expresses the divergence of a contravariant tensor field without a reference to the Christoffel symbols, $T^i\vert_i=\frac{1}{\sqrt{g}}\frac{\partial}{\partial x^i} (\sqrt{g}T^i)$, when applied to the inertialess Cauchy's equation of motion it yields
    \begin{equation}
        \frac{1}{\sqrt{g}}(\sqrt{g}\tau^{ij} \vec{g}_j)_{,i}+\rho  f^j\vec{g}_j=0,
    \end{equation}
    where $\tau^{ij}$ is the Cauchy stress tensor and $f^j$ are the contravariant components of volume forces. These stresses must be expressed both per unit area of a single reference surface and relative to base vectors that do not vary across the thickness of the sheet. The natural choice for this purpose is the midsurface and its intrinsic base vectors $\vec{a}_i$. If we use $\g_\lambda=\mu_\lambda^\alpha\A_\alpha$, $\g_3=\vec{n}$, the equilibrium equations can be rewritten in the following form \cite{GreenZerna}
    \begin{subequations}\label{eq:equilibrium}
    \begin{align}
        &\sigma^{\alpha\beta}\vert_{\alpha}-b^{\beta}_\alpha\sigma^{\alpha 3}+\sigma^{3\beta}_{,3}=-h\rho \mu^{\beta}_\lambda f^\lambda, 
        \quad\,\, \text{in-plane equilibrium} \label{eq:equilibrium_in_plane}
        \\
        &\sigma^{\alpha3}\vert_{\alpha}+b_{\alpha\beta}\sigma^{\alpha \beta}+\sigma^{33}_{,3}=-h\rho f^3, \label{eq:equilibrium_out_of_plane}
        \qquad \text{out-of-plane equilibrium}
    \end{align}
    \end{subequations}
    where 
    \begin{equation}
        \sigma^{i\lambda} = h\mu^\lambda_\alpha \tau^{i\alpha},\quad \text{and }\quad \sigma^{i3}=h\tau^{i3}
    \end{equation}
     are tangential and transverse stress tensor components referred to the midsurface. Notice that $\sigma^{ij}$ is not symmetric. Moreover, because of the symmetry of $\tau^{ij}$, it follows that $\sigma^{3\beta}=\mu ^\beta_\lambda \sigma^{\lambda 3}$.

\subsection{The cell cortex: bulk equations}
\paragraph{3D constitutive relations}
The cell cortex is often described as an active polymeric gel \cite{murrell2015forcing}. Polymer gels are cross-linked networks formed by linear or branched polymers. In the cell cortex, actin polymers are furthermore permanently renewed by polymerization/depolymerization, which releases elastic stresses over timescales of dozens of seconds~\cite{fritzsche2013analysis, Khalilgharibi:2019jt}. The cell cortex is therefore often described as a Maxwell viscoelastic material, responding elastically on short timescales and viscously on longer times, \cite{Kruse_EPJE_2005}. The cell cortex differs from physical gels in which it is an active material:  it is out-of-equilibrium, continuously consuming energy in the form of ATP.  Here, we work in the viscous regime of cortical behavior and start from viscous active nematic gel equations  \cite{Kruse_EPJE_2005, JoannyProst_HFSP_2009, Salbreux_PRL_2009}, which consider the stress to be the sum of a passive viscous term, and an active stress exerted by myosin molecular motors in the network $\vec{\tau} = \vec{\tau}_{pas}+\vec{\tau}_{act} $. The viscous response is modeled as a Newtonian fluid
    \begin{equation}
        \tau_{pas}^{ij} = -p g^{ij} +2\mu g^{ik}g^{jl}e_{kl},
    \end{equation}
    where $p$ is the hydrostatic pressure, $\mu$ is the shear viscosity of the actomyosin gel, and 
    \begin{equation}
        e_{ij}=\frac{1}{2}\left( \vec{g_i} \cdot \frac{\partial \vec{u}}{\partial \xi^j}+\vec{g_j}\cdot \frac{\partial\vec{u}}{\partial \xi^i}\right)
    \end{equation} 
    is the three-dimensional strain rate tensor, and $\vec{u}(\xi^{\alpha}, z, t)$ is the Lagrangian fluid velocity of the shell. 

\begin{figure}[ht]
\centering
\includegraphics[scale=1]{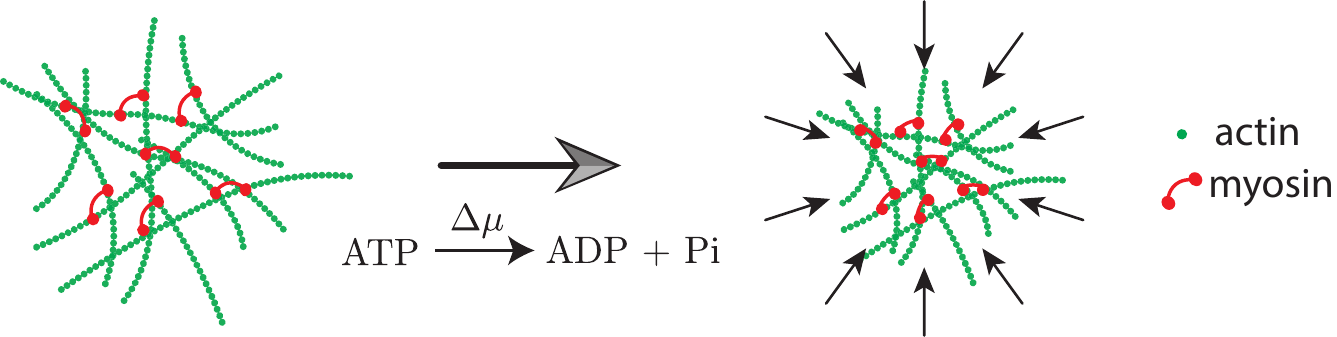}
\caption{Cortex as an active gel. The myosin transforms the energy of ATP hydrolysis into mechanical work.} \label{fig:contraction}
\end{figure}

\begin{figure}[ht]
\centering
\includegraphics[scale=1]{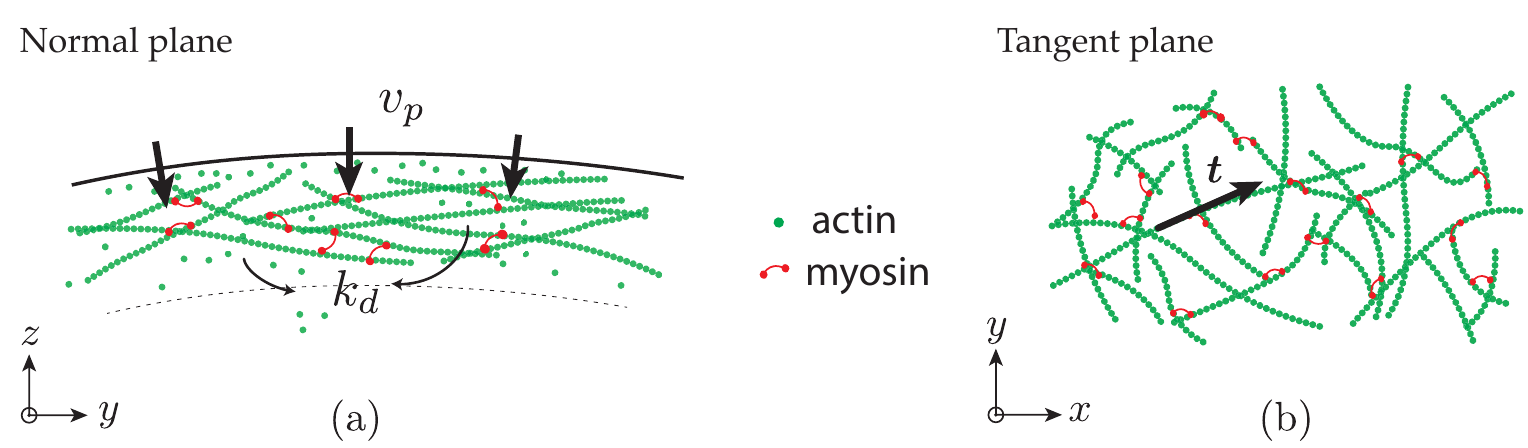}
\caption{(a) Visualization of a cross section of the cortex containing the normal to the midsurface. The actin filaments are mainly organized parallel to the midsurface. The addition of new actin monomers happens at the surface in contact with the plasma membrane, while the depolymerization happens uniformly across the cortex thickness. (b) In the tangent plane, the vector $\bm{t}$ represents the average orientation of an actin filament.} \label{fig:nematics}
\end{figure}

The cell cortex generates internal stress from the action of the motor protein myosin. Myosin motors can walk along actin filaments driven by the hydrolysis of ATP. When assembled into minifilaments and associated with several actin filaments, they induce internal stresses in the actin network (see Fig.~\ref{fig:contraction}).
While the detailed microscopic processes for the generation and renewal of motor-generated stress are complex \cite{lenz2012contractile}, they lead most generally to contractile stress in actomyosin networks \cite{Bendix:2008ima,murrell2015forcing}, an effect which is akin, in first approximation, to a negative pressure. The contractile activity is controlled spatiotemporally by cells by local phosphorylations of myosin chains through various biochemical pathways, such as the RhoA pathway \cite{AGARWAL2019150}.
As a result of motor activity or viscous flow, actin filaments may also acquire a structural organization within the cortex \cite{reymann2016cortical,Spira:2017}, that is generally measured by a director field $\boldsymbol{t}$ (see Fig.~\ref{fig:nematics}(b)) and described with a polar or nematic\footnote{In contrast to polar order, the nematic phase is described by the breaking of rotational symmetry, but not translational symmetry. In the nematic phase, the molecules are randomly polarized, but their long axes are oriented on the average along a particular direction specified by a unit vector $\bm{t}$, see Fig.~\ref{fig:nematics}} order parameter \cite{Kruse_EPJE_2005,Salbreux_PRL_2009}. While actin filaments turn over fast, the maintenance of a local orientation may be linked, experimentally, to crosslinking \cite{sobral2021plastin} or filament-guided de novo assembly \cite{li2021filament}. The active gel theory predicts an active stress contribution of the general form 
    \begin{equation}
    \label{eq:activestress}
        \tau_{act}^{ij} = \tilde{\zeta} \Delta \mu Q^{ij},
    \end{equation} 
where $\Delta \mu$ is the gain in chemical free energy for one ATP molecule hydrolysis, $\tilde{\zeta}>0$ is a measure of the local motor contractile activity \cite{Kruse_EPJE_2005}. At this point, we do not specify any particular form of the tensor $Q^{ij}$, which can represent nematic or polar order\footnote{In the nematic case, one can further divide the contractile stress in an isotropic and deviatoric components. This is done by simply decomposing $Q^{ij}$ such that
    \begin{equation}
        \tau_{act}^{ij} = \tilde{\zeta} \Delta \mu(\bar{\zeta} g^{ij} + \tilde{Q}^{ij}) = \underbrace{\zeta' \Delta \mu  g^{ij}}_{\text{Isotropic}} + \underbrace{\tilde{\zeta} \Delta \mu \tilde{Q}^{ij}}_{\text{Deviatoric}},
    \end{equation} 
The local orientation of actin filaments is measured by a traceless nematic tensor 
    $\tilde{Q}^{ij} = \left< t^i t^j - \frac{1}{3}\delta^{ij}\right>$,
    where $<>$ represents a local (mesoscopic) average over the orientation of the filaments in a elementary mesoscopic volume \cite{Prost_1993,doi2013soft}, and the orientation of each filament being characterized by a unit vector $\boldsymbol{t}$, see Fig.~\ref{fig:nematics}(b).}.
    Note that we neglect here the contribution of the conjugate field, a classical passive term in polar or nematic hydrodynamic theories, that reflect the restoring forces deriving from elastic energy penalizing gradients of filament orientations (the Frank-Oseen energy)\cite{Prost_1993}.

    The bulk constitutive relation is written in its contravariant form
    \begin{equation}\label{eq:ConstitutiveRelation}
        \tau^{ij}=-pg^{ij}+2\mu g^{ik}g^{jl}e_{kl}+\zeta  g^{ik}g^{jl}Q_{kl},
    \end{equation}
    where we introduced, to simplify notations, $\zeta \equiv \tilde{\zeta}\Delta \mu $: the magnitude of the actively generated stress in the cortex, which has to be positive to correspond to contractile stress. Note that we count positively the active stress and consider a positive phenomenological parameter $\tilde{\zeta} \geq 0$, taking a different sign convention than in \cite{Kruse_EPJE_2005}.

\paragraph{Local mass balance}
        
    The actomyosin cortex is under permanent turnover. The polymerization of new actin monomers happens mostly at the surface in contact with the plasma membrane at $z= T/2$, creating a flux of actin inwards the cell at a velocity $v_p = k_p\,a$, where $k_p$ is the rate of polymerization and $a$ the typical size of a monomer. The depolymerization of the actin may happen through different mechanisms \cite{fritzsche2013analysis}. Here we suppose that depolymerization is a bulk process that essentially happens at each free filament end, where monomers depolymerize at a rate $k_d$, see Fig.~\ref{fig:nematics}(a). Supposing actomyosin density uniform across the membrane, bulk mass balance reads 
    \begin{equation}\label{eq:mass_balance}
        \text{div } \vec{u}=-k_d+v_p\delta\left(z\!-\!\frac{T}{2}\right),
    \end{equation}
    where the delta function constrains polymerization from the upper shell surface boundary $z=T/2$. 

    \subsection{Balance of stress and torque resultants}  
    The fundamental dynamic quantities of interest for a thin shell theory are stress resultants and bending moments, which are integrals of the stress and torque across the sheet.  Therefore, we integrate the equations \eqref{eq:equilibrium} with respect to $z$ through the thickness of the shell and obtain
    \begin{equation}\label{eq:equilibrium_resultant}
    \begin{split}
        n^{\alpha\beta}\vert_\alpha-b^\beta_\alpha q^\alpha+\hat{P}^\beta=0,
        \\
        q^{\alpha}\vert_\alpha+b_{\alpha\beta}n^{\alpha\beta}+\hat{P}^3=0,
    \end{split}
    \end{equation}
    where \[ n^{\alpha\beta} \equiv \int_{-T/2}^{T/2}\sigma^{\alpha\beta}\dd z, \quad q^\alpha \equiv \int_{-T/2}^{T/2}\sigma^{\alpha3}\dd z,\] are the stress resultant tensor and transverse shear stress.
    $\hat{P}^\beta=\int_{-T/2}^{T/2} \mu^\beta_\lambda f^\lambda \rho h \, \dd z+F^\lambda_++F^\lambda_-$, and $\hat{P}^3=\int_{-T/2}^{T/2} f^3 \rho h \, \dd z+F^3_++F^3_-$ , with $ F^j_\pm=\Lambda_\pm\sigma^{ij}_\pm n^\pm_i,$ are the components of the stress applied in the direction $\A_j$ to the sheet outer surface, measured per unit area of the midsurface, and $\Lambda_\pm=(1+\frac{1}{4}g^{\alpha\beta}_\pm T_{,\alpha}T_{,\beta})^{1/2}$. The outward unit normal vectors are $\vec{n}^\pm=n_i^\pm\g^i_\pm$, with $n^\pm_\alpha=-\Lambda^{-1}_\pm \frac{1}{2}T_{,\alpha} $ and $n^\pm_3=\pm\Lambda^{-1}_\pm$. 

    To obtain the equation for the torque resultant we multiply \eqref{eq:equilibrium_in_plane} by $z$, then integrate across the sheet, 
    \begin{equation}\label{eq:moment}
        m^{\alpha\beta}\vert_{\alpha} + q^{\beta} = \mathcal{M}^\beta,
    \end{equation} 
    where \[ m^{\alpha\beta} \equiv - \int_{-T/2}^{T/2}\sigma^{\alpha\beta}z \,\dd z\] is the bending moment tensor, and $\mathcal{M}^\beta=\int_{-T/2}^{T/2}\mu^\beta_\lambda f^\lambda \rho h z \dd z+\frac{T}{2}(F^\beta_+-F^\beta_-)$ is the applied moment vector.

Next, we then introduce the \textit{symmetric} effective membrane stress tensor
    \begin{equation}\label{eq:N}
        N^{\alpha\beta}=n^{\alpha\beta}-b^{\beta}_\lambda m^{\alpha\lambda},
    \end{equation}
    and the \textit{symmetric} effective moment tensor as
    \begin{equation}\label{eq:M}
        M^{\alpha\beta}=\frac{1}{2}(m^{\alpha\beta}+m^{\beta\alpha}).
    \end{equation}
Then, we eliminate the shear stress resultant $q^\alpha$ using equation Eq.~\eqref{eq:moment}, and substitute \eqref{eq:N} and \eqref{eq:M} into \eqref{eq:equilibrium_resultant} to finally obtain~\citep{niordson2012shell}
    \begin{subequations}\label{eq:equilibrium_resultant_moment}
    \begin{align}
        N^{\alpha\beta}\vert_\alpha + 2 b^\beta_\gamma M^{\alpha\gamma}\vert_{\alpha} + M^{\gamma\alpha} b^\beta_\gamma \vert_\alpha+ P^\beta=0,\label{eq:equ_in_plane}
        \\
        M^{\alpha\beta}\vert_{\alpha\beta} - b_{\alpha\gamma} b^\gamma_\beta M^{\gamma\alpha} - b_{\alpha\beta}N^{\alpha\beta}-P^3=0,\label{eq:eq_out_plan}
    \end{align}
    \end{subequations}
    where $P^\beta=\hat{P}^{\beta}+b^\beta_\alpha \mathcal M^\alpha$, and $P^3=\hat{P}^3-\mathcal M^\alpha\vert_\alpha$.

    
    It is worth noting that the balance equations \eqref{eq:equilibrium_resultant_moment} are independent of the constitutive law for the shell: they represent universal force balance equations, valid for any 2-dimensional medium equipped with tension and bending moments, and simply expresses the conservation of momentum. Such decomposition is classical in the theory of elastic shells \cite{Green_1962, niordson2012shell}, and was also used to model the dynamics of viscous sheets \cite{Ribe_JFM_2002}. Moreover, since inertia has been neglected, time does not explicitly appear in the equilibrium equations or boundary conditions.

\section{Thin sheet constitutive equations} \label{sec:thinsheetequations}

    Because we are deriving equations of a shell theory based on stress resultants (Eq.~(\ref{eq:N}-\ref{eq:M}), constitutive equations for these stress resultants have to be formulated. These constitutive relations represent the stress field integrated over the shell thickness. Its formulation is not trivial since one must carefully consider the phenomenological three-dimensional constitutive equations for the visco-active material and for mass balance, introducing the active turnover effects into the two-dimensional constitutive relations.

    We now wish to exploit the slenderness of the cortex to obtain approximated "thin sheet" equations. Such an approach is valid when the slenderness parameter $\e = T_0/L \ll 1$. Then, we use asymptotic expansion in the small parameter $\varepsilon$ and in $z$ to express the effective stress $N^{\alpha\beta}$ and bending moment $M^{\alpha \beta}$ in terms of the midsurface velocity $\vec{U}\equiv \vec{u}(z \!=\! 0, t) = U^{\alpha} \vec{a}_\alpha +U^3\vec{n}$, in the local basis $\{\vec{a}_\alpha, \vec{n}\}$. More specifically, the constitutive equations will depend on the two tensors that describe the rate of deformation of the midsurface: the strain rate tensor $\Delta_{\alpha\beta}$, which is one half of the time derivative of the metric tensor $2\Delta_{\alpha\beta} \equiv \frac{d a_{\alpha\beta}}{d t}$, and the rate of change of the curvature tensor $\Omega_{\alpha\beta}$, which is defined as the time derivative of the curvature tensor $\Omega_{\alpha\beta} \equiv \frac{d b_{\alpha\beta}}{d t}$. These tensors can be explicitly written in terms of midsurface velocity $\vec U$ as \cite{Ribe_JFM_2002}
\begin{equation}\label{eq:Delta}
    \Delta_{\alpha\beta}=\frac{1}{2}(U_{\alpha\vert\beta}+U_{\beta\vert\alpha})-b_{\alpha\beta}U_3,
\end{equation}
    and 
\begin{equation}\label{eq:omega}
    \Omega_{\alpha\beta}= U_{3,\alpha\vert\beta}+b^\lambda_{\beta} U_{\lambda\vert\alpha}+b^\lambda_{\beta\vert\alpha} U_\lambda+b^\gamma_\beta U_{\gamma\vert\alpha}-b^\gamma_\beta b_{\alpha\gamma} U_3.
\end{equation}

    The deformation rate of the middle surface is described completely by the pair of symmetric tensors $\Delta_{\alpha\beta}$ and $\Omega_{\alpha\beta}$. It is important to note that the rate of tangential stretching depends not only on the rate of surface strain $(U_{\alpha\vert\beta}+U_{\beta\vert\alpha})/2$, but also on surface shape changes in the presence of curvature $-b_{\alpha\beta}U_3$.

\subsection{Non-dimensionalization and scaling analysis}

    The effective shear viscosity $\mu$ of the cortex is supposed to be fixed, and $P$ is a characteristic stress. Tangential coordinates scale as $x_\alpha = L \hat{x}_\alpha$, while the normal coordinate is $z = \e L  \hat z $.   Next, we define the dimensionless curvatures $\hat b_{\alpha\beta} = L b_{\alpha\beta}, \, \hat H =L H,\, \hat G =L^2 G $. Then, introduce the dimensionless volume element $h\equiv \sqrt{g/a}=1-2\hat{H}\hz\e+\hat{G}\hz^2\e^2,$ the dimensionless shift tensor $\mu^\beta_\gamma=\delta^\beta_\gamma-\e\hz \hat b^\beta_\gamma,$ and we approximate the contravariant metric tensor as, $g^{\alpha\beta}\approx a^{\alpha\beta}+2 \e\hz \hat b^{\alpha\beta}+\mathcal{O}(\e^2)$. 

    In theories of thin sheets, one often distinguishes between two main deformation modes that are highly dependent on the load applied to the system\footnote{In the context of elastic shells, it is well established that the deformation modes also depend on the geometry and boundary conditions of the structure. For detailed description of the asymptotic behaviors  we refer the reader to \cite{Chapelle_FEM_2011}}.  In the \textit{membrane} (or stretching-dominated) regime, the system responds mainly through stretching such that bending moments can be neglected. While in the \textit{inextensional} (or bending-dominated) regime, the dissipation issued from bending deformation rates becomes significant. These two regimes are described by different scalings for velocity and stress components \cite{Ribe_JFM_2001, Pfingstag_JFM_2011, Howell_EPAM_1996}. 

    In the membrane limit, the velocities and stresses in a viscous sheet scale as \cite{Fliert_JFM_1995}
    \begin{equation}\label{eq:scaling_membrane}
        \boldsymbol{u} \sim \frac{P L}{\mu \e}, \, p\sim \frac{P}{\e}.
    \end{equation}
    While in the inextensional (bending) regime, the appropriate scaling is \cite{buckmaster1975buckling, Ribe_JFM_2002}
    \begin{equation}\label{eq:scaling_bending}
        \boldsymbol{u} \sim \frac{P L}{\mu \e^3},\,  p\sim \frac{P}{\e^2}.
    \end{equation}

    In order to cover both deformation modes within our thin sheet model, we use the method of matched asymptotic expansions \cite{Fraenkel:1969}, which consists of finding approximate solutions separately in each limiting case, and then combining these approximations, giving an approximate constitutive relation that is valid for the whole range of intermediate behaviors.

\subsection{Asymptotic expansion: membrane limit}     
\label{app:asymptotic_membrane}

    We expand the variables $u_i$ and $p$ into a double power series in $\varepsilon=T_0/L$, and in the dimensionless normal coordinate $\hz=z/T_0$. The scaling \eqref{eq:scaling_membrane} indicate the appropriate expansion in the membrane limit
    \begin{equation}\label{eq:asymptoticmembrane}
        u_i = \frac{PL}{\mu \e}\sum_{m,n=0}\varepsilon^m\hat{z}^n u_i^{(mn)}, \quad
        p = \frac{P}{\e}\sum_{m,n=0}\varepsilon^m\hat{z}^n p^{(mn)}
    \end{equation}
    where $u_i ^{(mn)}$ and $p^{(mn)}$ are dimensionless functions of the curvilinear coordinates $x_1$ and $x_2$.
    
    We postulate that the active stress scales as the pressure, and the turnover scales as velocities (modulo a length for the depolymerization rate), which gives the following scalings for active parameters: the active stress $\zeta = \frac{P}{\e} \hat{\zeta} $, the polymerization velocity $v_p = \frac{PL}{\mu \e} \hat{v}_p$, and the depolymerization rate $k_d = \frac{P}{\mu \e} \hat{k}_d$.
    
    In the membrane limit, the bending moment $m^{\alpha\beta} \sim \mathcal{O}(\e^2)$, and at lowest order, $N^{\alpha\beta} = n^{\alpha\beta}$. 
    \begin{equation}
        N^{\alpha\beta}=\e L \int_{-\hat{T}/2}^{\hat{T}/2} (1-2\hat{H}\hz\e+\hat{G}\hz^2\e^2)(\delta^\beta_\gamma-\e\hz b^\beta_\gamma) (-pg^{\alpha\gamma}+2\mu g^{\alpha \lambda }g^{\gamma \delta}e_{\lambda \delta }+\zeta  g^{\alpha \lambda }g^{\gamma \delta}Q_{\lambda \delta }) \dd \hz.
    \end{equation}
    
    If the Lagrangian fluid velocity is decomposed in the local basis
    $
     \vec{u}(\xi^\alpha, z, t) = u^\alpha (\xi^\alpha, z, t) \vec{g}_\alpha + u_3(\xi^\alpha, z, t) \vec{n},
    $
    it follows that the strain rate tensor can be written as \cite{GreenZerna, Chapelle_FEM_2011}
        \begin{equation}\label{eq:strainrate}
        \begin{split}
        2e_{\alpha \beta}&=\mu^\lambda_\beta(u_{\lambda\vert\alpha}-b_{\lambda\alpha}u_3)+\mu^\lambda_\alpha(u_{\lambda\vert\beta}-b_{\lambda\beta}u_3), \\
        2e_{\alpha 3}&=u_{3,\alpha}+u_{\alpha,3}+b^\lambda_\alpha(u_\lambda-zu_{\lambda,3}), \\
        e_{33}&=u_{3,3}.
        \end{split}
    \end{equation}
One may remark that, at this point, no assumption was made on the kinematic, and our shell model allows for transverse shear and transverse normal deformations.
Now,  after applying \eqref{eq:asymptoticmembrane} in Eq.~\eqref{eq:strainrate}, and recalling that $\mu^\lambda_\beta = \delta_\beta^\lambda-\e\hz \hat{b}_\beta^\lambda$, we obtain the strain rate tensor
\begin{equation}\label{eq:strainrateexpansion}
\begin{split}
2e_{\alpha \beta} & = \frac{P}{\mu \e}\sum_{m,n=0}\varepsilon^m\hat{z}^n
\left[ u_{\alpha\vert\beta}^{(mn)}+u_{\beta\vert\alpha}^{(mn)}-2\hb_{\alpha\beta}u_3^{(mn)} \right.
 +\left.\e\hz(2\hb^\lambda_\beta\hb_{\lambda\alpha}u_3^{(mn)}-\hb^{\lambda}_\alpha u_{\lambda\vert\beta}^{(mn)}-\hb^{\lambda}_\beta u_{\lambda\vert\alpha}^{(mn)})\right].
\end{split}
\end{equation}

    Next, we substitute the expansion \eqref{eq:asymptoticmembrane} for the pressure, use \eqref{eq:strainrateexpansion} for the strain rate $e_{\lambda \delta}$, and use the approximation for the contravariant metric tensor $g^{\alpha\beta}\approx a^{\alpha\beta}+2 \e\hz b^{\alpha\beta}+\mathcal{O}(\e^2)$. 
    Then the leading term of the stress resultant is 
    \begin{equation}\label{eq:nab}
        \frac{N^{\alpha\beta}}{PL} = \hat{T} \left[2a^{\alpha\mu}a^{\beta\delta} \Delta^{(00)}_{\mu\delta} - a^{\alpha\beta} p^{(00)} \right] +  \hat{T} \zeta  a^{\alpha \mu} a^{\beta \delta} Q_{\mu \delta},
    \end{equation}
    where we introduced the dimensionless strain rate tensor $\Delta^{(00)}_{\alpha\beta}=\frac{1}{2}(u_{\alpha\vert\beta}^{(00)}+u_{\beta\vert\alpha}^{(00)})-\hb_{\alpha\beta} u_3^{(00)}$, which relates to the dimensional strain rate tensor as $\Delta_{\alpha\beta} = \frac{P}{\e \mu } \Delta^{(00)}_{\alpha\beta}$.

    To obtain a close relation in terms of velocities referred to the midsurface $u_i^{(0 0)}$, we need to find relations for the remaining coefficient $p^{(00)}$. These relations are found using the expansion \eqref{eq:asymptoticmembrane} into the mass balance equation \eqref{eq:mass_balance}, and boundary condition (Eq.~\eqref{eq:bc} in App.~\ref{app:bc}); and then equate to zero the factors in the expansion proportional to the same power in $\e$ and $\hat z$ until the leading term $u_i^{(0 0)}$ can be fully identified. This will lead to a set of algebraic equations that can be solved sequentially.

The continuity of normal stress implies, from \eqref{eq:bc}, $\tau^{33}_\pm = P^3_{\pm}$, see App.~\ref{app:bc}. Then, with the expansions \eqref{eq:asymptoticmembrane},
    \begin{equation*}
    \begin{split}
        \e^{-1}\sum_{m,n=0} \varepsilon^m(\pm\hat{T}/2)^n 
        \left[-p^{(mn)}+  2 (n+1)u_3^{(m+1n+1)} \right] +\e^{-1}\hat \zeta Q_{33} = P^3_\pm.
    \end{split}
    \end{equation*}
    Considering the lowest order term corresponding to $m=0,n=0$, we have 
    \begin{equation}\label{eq:p00_active}
        p^{(00)} = 2u_3^{(11)}+ \hat \zeta Q_{33}.
    \end{equation}

Next, from $ \text{div}\,\vec{u} = \vec{g}^i\vec{u}_{,i}$, we rewrite the mass balance \eqref{eq:mass_balance} as
    \begin{equation}\label{eq:mass_conserv}
        (hu_3)_{,3}+[a^{\alpha\beta}+z(b^{\alpha\beta}-2Ha^{\alpha\beta})]u_{\alpha\vert\beta}=-k_d h + v_p \delta\left(z\!-\!\frac{T}{2}\right) h,
    \end{equation}
    where $h$ is given by \eqref{eq:h}. Then, we introduce the expansions \eqref{eq:asymptoticmembrane},
    \begin{equation}
    \begin{split}
        \sum_{m,n=0}\varepsilon^m\hat{z}^n
        \left[   (n+1)u_3^{(m+1n+1)} +a^{\alpha\beta} u_{\alpha\vert\beta}^{(mn)}- 2\hat{H}u_3^{(mn)}  + \mathcal{O}(\e)  \right]
        =  -\hat k_d+\hat v_p\delta(\hat z-\hat T/2) 
        + \mathcal{O}(\e).
    \end{split}
    \end{equation}
    Anticipating that we only needed the lowest order terms, the expansion was truncated to $\mathcal{O}(\e)$. The zeroth order term, $(\e\hz)^0$, is achieved when $m=0$, and $n=0$, and gives
    \begin{equation}
        u_3^{(11)}=-a^{\alpha\beta}\Delta_{\alpha\beta}^{(00)}-\hat k_d+\frac{\hat v_p}{\hat T} 
    \end{equation}

    From \eqref{eq:p00_active}, we can further obtain
    \begin{equation}\label{eq:p00}
        p^{(00)}=-2a^{\alpha\beta}\Delta_{\alpha\beta}^{(00)}-2\hat k_d +\frac{2\hat v_p}{\hat T} + \hat  \zeta Q_{33}.
    \end{equation}


    By substituting  Eq.~\eqref{eq:p00} into Eq.~\eqref{eq:nab}, and rewriting the terms in its dimensional form we obtain the symmetric stress resultant
    \begin{equation}\label{eq:Nab_membrane}
        N^{\alpha\beta} = 4 \mu T \mathcal{A}^{\alpha\beta\mu\delta}\Delta_{\mu\delta}+2\mu T a^{\alpha\beta}\left(k_d-\frac{v_p}{T}\right)+T\mathcal{Q}^{\alpha\beta}\zeta ,
    \end{equation}
    where $\mathcal{A}^{\alpha\beta\mu\delta}=\frac{1}{4}(a^{\alpha\mu}a^{\beta\delta}+a^{\alpha\delta}a^{\beta\mu})+\frac{1}{2} a^{\alpha\beta}a^{\mu\delta}$,  $\mathcal{Q}^{\alpha\beta}=a^{\alpha\mu}a^{\beta\delta} Q_{\mu \delta}-a^{\alpha\beta}Q_{33}$.

    We find therefore that active tension contributions are generated here not only by contractile motor activity $T\mathcal{Q}^{\alpha\beta}\zeta$, as proposed earlier in axisymmetric geometry \cite{Mietke_PNAS_2019, Turlier_BioPhys_2014} or in general three dimensional settings \cite{Arroyo_JFM_2019}, but also through active turnover $2\mu T a^{\alpha\beta}(k_d-v_p/T)$. This new active tension term due to turnover, which was overlooked so far, can lead to either a contractile or extensile stress and constitutes an important theoretical prediction of our manuscript.


    \subsection{Asymptotic expansion: Inextensional limit}
    
    The appropriate scaling for the inextensional regime \eqref{eq:scaling_bending} gives the following expansion
    \begin{equation}\label{eq:asymptoticbending}
        u_i=\frac{PL}{\mu \e^3}\sum_{m,n=0}\varepsilon^m\hat{z}^n u_i^{(mn)},
        \quad
        p=\frac{P}{\e^2}\sum_{m,n=0}\varepsilon^m\hat{z}^n p^{(mn)},
    \end{equation}
and as in the membrane regime, we hypothesize that active sources of stress scale similarly to the sheet dynamic variables:  $\zeta = \frac{P}{\e^2} \hat{\zeta} $, $v_p = \frac{PL}{\mu \e^3} \hat{v}_p$, and $k_d = \frac{P}{\mu \e^3} \hat{k}_d$.
    We then use a similar approach to obtain leading-order expressions for the dimensional (symmetric) stress and bending resultants $N^{\alpha\beta}$, $M^{\alpha\beta}$.
    For the sake of the presentation, we defer detailed calculations to Appendix \ref{app:Asymptotic_bending}.
    
    We find for the tension resultant
\begin{equation}
\begin{split}
    N^{\alpha\beta} & =\, 4 \mu T \mathcal{A}^{\alpha\beta\mu\delta}\Delta_{\mu\delta} + T \zeta  a^{\alpha \mu} a^{\beta \delta} Q_{\mu \delta}
    \\
    & +  \frac{T^3}{12}\left[\mu C^{\alpha\beta\mu\delta}\Omega_{\mu\delta} + \mu B^{\alpha\beta} \left( k_d - \frac{ v_p}{ T}\right)
     + \zeta \mathcal{D}_s^{\alpha\beta\mu\delta}Q_{\mu\delta} \right] ,
\end{split}
\end{equation}
where   $\mathcal{C}^{\alpha\beta\mu\delta}=8H\mathcal{A}^{\alpha\beta\mu\delta}-a^{\alpha\beta}b^{\mu\delta}-5b^{\alpha\beta}a^{\mu\delta}-\frac{3}{2}(a^{\alpha\mu}b^{\beta\delta}+a^{\alpha\delta} b^{\beta\mu}+a^{\beta\mu}b^{\alpha\delta}+a^{\beta\delta} b^{\alpha\mu})$, $B^{\alpha\beta} = 2 (G -H^2) a^{\alpha\beta} +6 H b^{\alpha\beta} +10 b^\alpha_\mu b^{\beta\mu}$, and $\mathcal{D}_s^{\alpha\beta\mu\delta} = G a^{\alpha\mu}a^{\beta\delta} - 4 H b^{\alpha\mu}a^{\beta\delta} - b^{\alpha\mu} b^{\beta\delta} -3 a^{\alpha\mu} b^{\delta\gamma}  b^\beta_\gamma$.

And for the bending moment
\begin{equation}
\begin{split}
    M^{\alpha\beta} =
    \mu \frac{T^3}{3}\mathcal{A}^{\alpha\beta\mu\delta}\Omega_{\mu\delta}
    + \mu \frac{T^3}{12}(6H a^{\alpha\beta}
    + 4b^{\alpha\beta})\left(k_d - \frac{v_p}{T}\right)
    + \frac{T^3}{12}\zeta \mathcal{D}_b^{\alpha\beta\mu\delta} Q_{\mu\delta},
\end{split}
\end{equation}
where  $\mathcal{D}_b^{\alpha\beta\mu\delta}=H(a^{\alpha\mu}a^{\beta\delta}+a^{\beta\mu}a^{\alpha\delta})+a^{\alpha\beta}\hb^{\mu\delta}-\frac{3}{4}(a^{\alpha\mu}\hb^{\beta\delta}+a^{\alpha\delta}\hb^{\beta\mu}+a^{\beta\mu}\hb^{\alpha\delta}+a^{\beta\delta}\hb^{\alpha\mu})$.

Note that in this regime the active turnover stress appear only in the terms proportional to $T^3$ in the membrane stress, and in the bending moment.

\subsection{Composite equations}
The composite equations for $N^{\alpha\beta}$ and $M^{\alpha\beta}$ are obtained by gathering together the different terms appearing in the membrane and inextensional limits. The approximate constitutive relation obtained by "matching" the terms is valid for the whole range of intermediate behaviors. The ensuing complete system is then,
\begin{equation}\label{eq:N_composite}
\begin{split}
    N^{\alpha\beta} = & 4 \mu T \mathcal{A}^{\alpha\beta\mu\delta}\Delta_{\mu\delta}+2\mu T a^{\alpha\beta}\left(k_d-\frac{v_p}{T}\right)+T\mathcal{Q}^{\alpha\beta}\zeta 
    \\
    & + \frac{T^3}{12}\left[\mu C^{\alpha\beta\mu\delta}\Omega_{\mu\delta} + \mu B^{\alpha\beta} \left( k_d - \frac{ v_p}{ T}\right)
     + \zeta \mathcal{D}_s^{\alpha\beta\mu\delta}Q_{\mu\delta} \right] ,
\end{split}
\end{equation}
and 
\begin{equation}\label{eq:M_composite}
\begin{split}
    M^{\alpha\beta} =
    \mu \frac{T^3}{3}\mathcal{A}^{\alpha\beta\mu\delta}\Omega_{\mu\delta}
    + \mu \frac{T^3}{12}(6H a^{\alpha\beta}
    + 4b^{\alpha\beta})\left(k_d - \frac{v_p}{T}\right)
    + \frac{T^3}{12}\zeta \mathcal{D}_b^{\alpha\beta\mu\delta} Q_{\mu\delta},
\end{split}
\end{equation}

In the absence of active terms, our constitutive relations are identical to the ones obtained in \cite{Ribe_JFM_2002} for a viscous sheet without couplings between bending and stretching.  
Moreover, we note that the absence of transverse shear deformation at the lowest order in the asymptotic expansion induces a Kirchhoff-Love type of deformation\footnote{The Kirchhoff-Love kinematical assumption states that the material line orthogonal to the midsurface in the undeformed configuration remains straight, unstretched, and orthogonal to the midsurface after deformation. In our particular case, the condition that the normal is unstretched is relaxed since we allow the shell to vary its thickness.}. For an initially curved elastic shell, Kirchhoff-Love deformations lead to a theory generally named after Koiter \cite{Koiter_1970}. This can be readily seen since the passive parts of the resultants \eqref{eq:N_composite} and \eqref{eq:M_composite} are, modulo a time derivative, similar to the incompressible limit of Koiter's constitutive equations for elastic shells \cite{Koiter_1970}, as one may expect from the Stokes-Rayleigh analogy\footnote{The Stokes-Rayleigh analogy addresses the similarities between the formulation of viscous and elastic problems. It states that the equations for slow viscous flow derive from the equations of linear elasticity when velocities replace positions and strain rates replace strains \cite{Stokes_1845, Rayleigh_1945}.}. 

\section{Variational formulation: weak form and Rayleigh potential}
\label{sec:variationalform}
The weak form is obtained by multiplying the balance equation \eqref{eq:equilibrium_resultant_moment} by a test function, then integrating over the region in space occupied by the shell,
\begin{subequations}\label{eq:wf1}
    \begin{align}
        \iint_{\mathcal{S}}\left(N^{\alpha\beta}\vert_\alpha + 2 b^\beta_\gamma M^{\alpha\gamma}\vert_{\alpha} + M^{\gamma\alpha} b^\beta_\gamma \vert_\alpha+ P^\beta \right)\delta U_{\beta} \dd A = 0,\label{eq:wfa}
        \\
        \iint_{\mathcal{S}}\left(M^{\alpha\beta}\vert_{\alpha\beta} - b_{\alpha\gamma} b^\gamma_\beta M^{\gamma\alpha} - b_{\alpha\beta}N^{\alpha\beta}-P^3\right) \delta U_3 \dd A = 
        0,\label{eq:wfb}
    \end{align}
\end{subequations}
where $\delta U_\beta(x^1,x^2)$, and $\delta U_3(x^1,x^2)$ are a set of "virtual velocities" (test functions). Using integration by parts, the divergence theorem for curved surfaces, and neglecting the work terms from the surface edges, the variational (weak) form is written as \cite{niordson2012shell}
    \begin{equation}\label{eq:weak_form}
        \iint_{\mathcal{S}}\left(N^{\alpha\beta}\delta\Delta_{\alpha\beta}+M^{\alpha\beta}\delta\Omega_{\alpha\beta} \right) \dd A =\iint_{\mathcal{S}}(P^\beta\delta U_\beta+P^3\delta U_3) \dd A.
    \end{equation}
    where $N^{\alpha\beta}$, and $M^{\alpha\beta}$ are defined in \eqref{eq:N}, and \eqref{eq:M}, and where 
     $\delta\Delta_{\alpha\beta}=\frac{1}{2}(\delta U_{\alpha\vert\beta}+\delta U_{\beta\vert\alpha})-b_{\alpha\beta}\delta U_3$, and $\delta \Omega_{\alpha\beta}=\delta U_{3, \alpha\vert\beta}+b^\lambda_\alpha \vert_\beta\delta U_\lambda+b^\lambda_\alpha\delta U_{\lambda\vert\beta}+b^\lambda_\beta \delta U_{\lambda\vert\alpha}-b^\lambda_\beta b_{\alpha\lambda}\delta U_3$. 

    The left hand side of the weak formulation \eqref{eq:weak_form} might be interpreted as internal virtual power, since we identified $\delta \Delta_{\alpha\beta}$ and $\delta\Omega_{\alpha\beta}$ as measures of strain rate increments. The terms in the right hand side of \eqref{eq:weak_form} can be interpreted as the virtual power of external loads.

\paragraph{The Rayleigh potential}
    Starting from the weak form \eqref{eq:weak_form}, if the set of virtual velocities are chosen to be the real velocities $\delta U_i = U_i$, one can identify the resulting variational function as a Rayleigh potential \cite{strutt1871some,Minguzzi_2015}, $\mathcal{R}$, which is the sum of the (passive) power dissipated by viscous forces $\mathcal{D}$, and the power input of active terms $\mathcal{P}$. 

    In the absence of external forces, the Rayleigh potential is
    \begin{equation}\label{eq:Rayleigh}
    \begin{split}
        \mathcal{R}\big(\boldsymbol{U},\boldsymbol{x}\big) &  = \iint_{\mathcal{S}}\left( N^{\alpha\beta}\Delta_{\alpha\beta}+M^{\alpha\beta}\Omega_{\alpha\beta} \right) \dd A 
        \\
        & = \overbrace{
        \iint_{\mathcal{S}} \Big[ 4 \mu T \mathcal{A}^{\alpha\beta\gamma\delta} \Delta_{\alpha\beta} \Delta_{\gamma\delta}  
         + \mu \frac{T^3}{3}  \mathcal{A}^{\alpha\beta\gamma\delta} \Omega_{\alpha\beta} \Omega_{\gamma\delta} +
         \mu \frac{T^3}{12}  \mathcal{C}^{\alpha\beta\gamma\delta} \Delta_{\alpha\beta} \Omega_{\gamma\delta} \Big] \dd A}^{\mathcal{D}}
         \\
         & +\overbrace{\iint_{\mathcal{S}} \Big[ \mu \left(T a^{\alpha\beta} +\frac{T^3}{12} B^{\alpha\beta} \right) \left( k_d -\frac{v_p}{T}\right)\Delta_{\alpha\beta} + \mu \frac{T^3}{12}(6H a^{\alpha\beta}
    + 4b^{\alpha\beta})\left(k_d - \frac{v_p}{T}\right)\Omega_{\alpha\beta}}^{\mathcal{P}} \\
         & \qquad \quad + \frac{1}{2} T \mathcal{Q}^{\alpha\beta} \zeta \Delta_{\alpha\beta}  + \frac{T^3}{12}\zeta Q_{\mu\delta} (\mathcal{D}_s^{\alpha\beta\mu\delta} \Delta_{\alpha\beta}+\mathcal{D}_b^{\alpha\beta\mu\delta}\Omega_{\alpha\beta})  \Big] \dd A
         \\
         & = \mathcal{D} +\mathcal{P}
    \end{split}
    \end{equation}
    
    The Rayleigh potential plays a similar role for viscous problems as the energy potential in elasticity, and allows to calculate the midsurface velocity $\boldsymbol{U}$ as its minimizer 
    \begin{equation}
        \mathcal{\boldsymbol{U}} = \argmin_{\boldsymbol{U}'}\mathcal{R}\big(\boldsymbol{U}',\boldsymbol{x}\big) 
    \end{equation}
    The midsurface geometry may then be deduced from the current velocity as $\frac{d\boldsymbol{x}}{dt}=\boldsymbol{U}$. In a sense, this simple variational form expresses here simply the fact that viscous flows minimize dissipation \cite{happel2012low}. A generalization of such approach for dissipative systems, to include other sources of dissipation but also reactive forces deriving from a free energy, is often named Onsager variational principle in soft-matter physics \cite{doi2011onsager,arroyo2018onsager}.
    
\section{Evolution of the sheet thickness}
\label{sec:thicknessevolution}
    To obtain the equation of evolution for the sheet thickness, one may integrate the equation \eqref{eq:mass_conserv}, as performed in \cite{Ribe_JFM_2002}. Alternatively, we derive here mass balance directly at the level of the shell.
    Consider a small volume of the sheet $\dd V = \int_{-T/2}^{T/2} \dd z \, h(z) \,\dd S$, where $h(z) = 1-2 Hz +G z^2$, and $\dd S(t) = \sqrt{a(t)} \,\dd \xi^1\dd \xi^2$ is a infinitesimal area of the sheet. For constant density $\rho$, the conservation of mass can be written as
    \begin{equation}
        \frac{d}{d t}\left( \int_{-T/2}^{T/2}  \, h(z) \,\dd  S \,\dd z  \right)  = \int_{-T/2}^{T/2} v_{p}\delta\left(z\!-\!\frac{T}{2}\right) \, h(z) \,\dd S \,\dd z - k_d \int_{-T/2}^{T/2}  \, h(z) \,\dd S \,\dd z.
    \end{equation}
    
    In the left hand side, we use Leibniz's integral rule to bring the derivative inside the integral, and since the integration limits $\pm T/2$ are time-dependent, this will give terms in $\partial T/\partial t$. Then,  using the fact that $ \frac{d }{dt}a = \frac{d }{dt} (\det a_{\alpha\beta})=\det a_{\alpha\beta}\, a^{\alpha\beta} \frac{d a_{\alpha\beta}}{dt}$, it follows from \eqref{eq:Delta}
    \begin{equation}
        \left(1+\frac{G T^2}{4}\right) \,\frac{\partial T}{\partial t}\,\dd S  +  T\left( 1+\frac{G T^2}{12}\right)a^{\alpha\beta} \Delta_{\alpha\beta}\,\dd S =  v_{p} \left(1- H T +\frac{GT^2}{4}\right)\,\dd S  - k_d T\left(1+\frac{GT^2}{12}\right) \,\dd S.
    \end{equation}
    Given $GT^2 \sim \mathcal{O}(\e^2)\ll 1$, the equation for the time evolution of the thickness becomes, at first order in $\e$
    \begin{equation}\label{eq:thickness}
        \frac{\partial T}{\partial t}=-T a^{\alpha\beta}\Delta_{\alpha\beta}-T k_d + v_p (1-HT ) + \mathcal{O}(\e^2).
    \end{equation}
        
    For an incompressible passive fluid, the equivalent equation would be simply $\frac{\partial T}{\partial t}=-T a^{\alpha\beta}\Delta_{\alpha\beta}$, which states that rate of change of the thickness is proportional to minus the shell stretching rate $a^{\alpha\beta}\Delta_{\alpha\beta}$, as found in \cite{buckmaster1975buckling,Ribe_JFM_2002}

\section{Summary of the main equations}
\label{sec:summaryModel}
\begin{tcolorbox}[]
Solve for the midsurface velocity $\bm{U}$ that minimizes the Rayleighian,
\begin{equation}
    \mathcal{\boldsymbol{U}} = \argmin_{\boldsymbol{U}'}\mathcal{R}\big(\boldsymbol{U}',\boldsymbol{x}\big),
\end{equation}
with
\begin{equation}
\begin{split}
    \mathcal{R}\big(\boldsymbol{U},\boldsymbol{x}\big)  = \iint_{\mathcal{S}}\left( N^{\alpha\beta}\Delta_{\alpha\beta}+M^{\alpha\beta}\Omega_{\alpha\beta} \right) \dd A,
\end{split}
\end{equation}
where the \textit{thin shell} constitutive equations are
\begin{equation}
\begin{split}
    N^{\alpha\beta} = & 4 \mu T \mathcal{A}^{\alpha\beta\mu\delta}\Delta_{\mu\delta}+2\mu T a^{\alpha\beta}\left(k_d-\frac{v_p}{T}\right)+T\mathcal{Q}^{\alpha\beta}\zeta 
    \\
    & + \frac{T^3}{12}\left[\mu C^{\alpha\beta\mu\delta}\Omega_{\mu\delta} + \mu B^{\alpha\beta} \left( k_d - \frac{ v_p}{ T}\right)
     + \zeta \mathcal{D}_s^{\alpha\beta\mu\delta}Q_{\mu\delta} \right] ,
\end{split}
\end{equation}
and 
\begin{equation}
\begin{split}
    M^{\alpha\beta} =
    \mu \frac{T^3}{3}\mathcal{A}^{\alpha\beta\mu\delta}\Omega_{\mu\delta}
    + \mu \frac{T^3}{12}(6H a^{\alpha\beta}
    + 4b^{\alpha\beta})\left(k_d - \frac{v_p}{T}\right)
    + \frac{T^3}{12}\zeta \mathcal{D}_b^{\alpha\beta\mu\delta} Q_{\mu\delta}.
\end{split}
\end{equation}

The midsurface velocity is implicitly written in terms of rate of deformation measures, namely the strain rate tensor 
\begin{equation}
    \Delta_{\alpha\beta}=\frac{1}{2}(U_{\alpha\vert\beta}+U_{\beta\vert\alpha})-b_{\alpha\beta}U_3,
\end{equation}
    and rate of change of the curvature tensor
\begin{equation}
    \Omega_{\alpha\beta}= U_{3,\alpha\vert\beta}+b^\lambda_{\beta} U_{\lambda\vert\alpha}+b^\lambda_{\beta\vert\alpha} U_\lambda+b^\gamma_\beta U_{\gamma\vert\alpha}-b^\gamma_\beta b_{\alpha\gamma} U_3.
\end{equation}

The geometry is then solved in a Lagrangian manner, the midsurface geometry position $\bm{x}$ is deduced from the current velocity
\begin{equation}
    \frac{d \bm{x}}{d t} = \bm{U},
\end{equation}
and the thickness spatio-temporal evolution is described by the following equation
\begin{equation}
    \frac{d T}{d t}=-T a^{\alpha\beta}\Delta_{\alpha\beta}-T k_d + v_p (1-HT ).
\end{equation}

\end{tcolorbox}

\section{Numerical approach}
\label{sec:numericalscheme}
\subsection{Preliminaries}
We discuss here some aspects regarding the finite-element solution scheme used for the numerical simulations. Note, however, that computational aspects of the viscous thin shell theory are not the primary purpose of the present work. The proposed implementation will therefore be limited to simple geometrical settings which excludes non-manifold surfaces for instance. For more details on the practical numerical implementation aspects, we refer to the documentation of the source code released with the paper \cite{githubFENICS}.
    
For viscous membranes, a few numerical implementations have been recently proposed based either using finite element models \cite{reuther2020numerical,Arroyo_JFM_2019} or discrete geometric approaches (computer graphics) \cite{grinspun2003discrete, batty2012discrete}. Yet, no numerical work has been dedicated to implementing viscous thin shells in generalized coordinates to our best knowledge. Let us first point out that thin shell theories based on the Kirchhoff-Love kinematics are notoriously difficult to implement due to the system being of fourth-order type, which requires $C^1$-continuity of the approximating finite element discretization fields to avoid "kinks," leading to infinite bending energies. 
Since this cannot be achieved with standard $\mathcal{C}^0$ finite elements, one has to resort to either dedicated or complex finite elements \cite{Cirak:2000cb, Cirak:2001iv, Chapelle_FEM_2011} or isogeometric analysis approaches \cite{kiendl2009isogeometric}. 
To circumvent the difficulty of handling $\mathcal{C}^1$-conforming elements, one may also relax the underlying thin shell hypothesis and solve an equivalent shearable shell model \cite{naghdi1973theory} which requires only $\mathcal{C}^0$ continuity. The use of such a \textit{shearable} (thick shell) instead of \textit{unshearable} (thin shell) kinematics must, however, consistently converge to the latter in the thin shell limit $\e \to 0$. In such thick (or Naghdi-type) shell models, the shell cross-section does not remain normal to the shell surface any more. Additional degrees of freedom are introduced in the form of \textit{inextensible directors}. These inextensible directors are allowed to deviate from the surface normal during the deformation, hence accounting for transversal shearing of the shell across the thickness. More simple $\mathcal{C}^0$ elements can then be considered in the case of thick shell kinematics for the both displacement/velocity and director fields. However, shear locking issues are prone to arise in the thin shell limit, requiring the use of dedicated numerical techniques such as mixed methods to circumvent the locking issue \cite{bathe2000evaluation, Hale_FEniCS-Shellls_2018}.
    
Our simulations were conducted along these lines, by relaxing the thin shell hypothesis in the numerical implementation by considering a director field $\boldsymbol{d}$ which, contrary to $\boldsymbol{n}$, is not necessarily normal to the shell midsurface, see Fig.~\ref{fig:Sherable_shell}. Strain measure expressions need then be adapted with $\boldsymbol{d}$ instead of $\boldsymbol{n}$.
Since the shell cortex is, in practice, extremely thin, we expect to recover the theoretical thin shell limit derived above by adding a penalization for the deviation between the normal $\boldsymbol{n}$ and the shell director $\boldsymbol{d}$ in the dissipation potential (see Eq.~\eqref{eq:phi}).


\subsection{Discrete shell geometry}
The discrete shell reference surface, denoted by $\boldsymbol{X}$, is assumed to be made of an assembly of flat triangular elements and $\boldsymbol{d}^0=\boldsymbol{n}^0$ denotes the reference shell director which we assume to initially coincide with the reference shell normal. The reference metric tensor and curvature tensors are defined respectively as $a_{\alpha\beta}^0 = \frac{1}{2} \big(\boldsymbol{X}_{,\alpha}  \boldsymbol{X}_{,\beta} + \boldsymbol{X}_{,\beta}  \boldsymbol{X}_{,\alpha} \big)$ and $b_{\alpha\beta}^0 = -\frac{1}{2} \big(\boldsymbol{X}_{,\alpha} \bm{d}^0_{,\beta} + \boldsymbol{X}_{,\beta} \bm{d}^0_{,\alpha} \big)$ \cite{Hale_FEniCS-Shellls_2018}. 
In the current configuration, the metric and curvature tensors are similarly given by $a_{\alpha\beta} = \frac{1}{2} \big(\boldsymbol{x}_{,\alpha}  \boldsymbol{x}_{,\beta} + \boldsymbol{x}_{,\beta}  \boldsymbol{x}_{,\alpha} \big)$ and $b_{\alpha\beta} = -\frac{1}{2}\big(\boldsymbol{x}_{,\alpha} \bm{d}_{,\beta} + \boldsymbol{x}_{,\beta} \bm{d}_{,\alpha}\big)$.

\begin{figure}[ht]
\centering
\includegraphics[scale=.95]{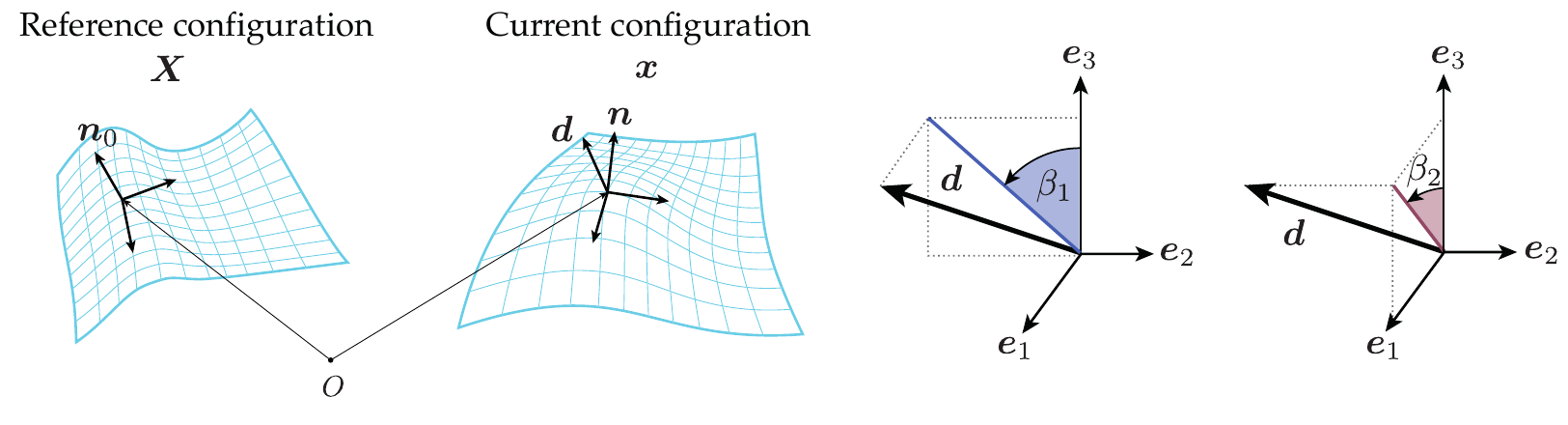}
\caption{Kinematic of the \textit{shearable} shell model. Left: generic point in the reference and current configurations of the shell midsurface. Right: Parametrization of the inextensible director $\bm{d}$ in terms of two independent Euler angles $(\beta_1,\beta_2)$.} \label{fig:Sherable_shell}
\end{figure}

Following \cite{Hale_FEniCS-Shellls_2018}, we assume the shell director to be unstretchable ($\|\boldsymbol{d}\|=1$) so that we parametrize the unit director $\boldsymbol{d}$ by two angles $(\beta_1,\beta_2)$ (see Fig.~\ref{fig:Sherable_shell} right), giving rise a so-called five-parameter shell model: \cite{betsch1998parametrization}
    \begin{equation}
      \boldsymbol{d}(\beta_1,\beta_2) =(\cos \beta_1 \sin \beta_2, -\sin \beta_1, \cos \beta_1 \cos \beta_2).
    \label{eq:director-parametrization}
    \end{equation}

 Note that this simple parametrization limits our implementation to non-closed manifold surfaces. We implement an updated Lagrangian formulation, in which the discrete equations are formulated in the current configuration, which is assumed to be the new reference configuration. In this setting, we replace the definitions of the strain rate tensor $\Delta_{\alpha\beta} = \frac{1}{2} \frac{d a_{\alpha\beta}}{dt}$ and rate of change of curvature tensor $\Omega_{\alpha\beta} = \frac{d b_{\alpha\beta}}{dt} $ with their linearized expressions for the rate of deformation \cite{Hale_FEniCS-Shellls_2018}:
    
    
    \begin{equation}\label{eq:discr-Delta_ind}
    \begin{split}
        \Delta_{\alpha\beta} & = \frac{1}{2}\big[ \bm{X}_{,\alpha} \bm{U}_{,\beta} + \bm{X}_{,\beta} \bm{U}_{,\alpha}\big],
    \end{split}
    \end{equation}
    and
    \begin{equation}
        \Omega_{\alpha\beta} = -\frac{1}{2}\left(\boldsymbol{U}_{,\alpha} \boldsymbol{d}^0_{,\beta} + \boldsymbol{U}_{,\beta}\boldsymbol{d}^0_{,\alpha} + \bm{X}_{,\alpha}  \boldsymbol{\omega}_{,\beta}  + \bm{X}_{,\beta } \boldsymbol{\omega}_{,\alpha} \right)
    \label{eq:discr-Omega_ind}
    \end{equation}
    where $\boldsymbol{\omega} \equiv \dot{\boldsymbol{d}} = \frac{d\boldsymbol{d}}{dt}$ is the rate of change of director, and $U = \frac{d \bm{x}}{dt}$. 
    
    In addition, we must also consider the rate of shear strain,
    \begin{equation}
        \gamma_\alpha =  \boldsymbol{\omega} \bm{X}_{,\alpha} + \boldsymbol{d}^0\boldsymbol{U}_{,\alpha} .
    \label{eq:discr-gamma_ind}
    \end{equation}
    When $\gamma_\alpha = 0$, we recover that $\omega = (U_{3,\alpha} + b^\lambda_\alpha U_\lambda)\bm{a}^\alpha$, then substituting into Eq.~\eqref{eq:discr-Omega_ind} we should retrieve exactly Eq.~\eqref{eq:omega}, which corresponds to the unshearable shell dynamics\footnote{We use the fact that $\bm{U}_{,\alpha} = (U_{\lambda|\alpha} - b_{\lambda\alpha}U_3)\bm{a}^\lambda + (U_{3,\alpha} + b^\lambda_\alpha U_\lambda)\bm{n}$, which comes from the decomposition of the velocity $U$ in the local frame and the use of the Gauss-Weingarten relations}.
    
    As a consequence, the new Rayleigh potential density is modified to add a novel dissipation contribution penalizing shear:
    \begin{equation}\label{eq:phi}
        \phi(\boldsymbol{U}, \boldsymbol{d}) =  N^{\alpha\beta} \Delta_{\alpha\beta} + M^{\alpha \beta}\Omega_{\alpha\beta}  +k_{s,\text{pen}} \mu T a^{\alpha\beta} \gamma_{\alpha} \gamma_{\beta},
    \end{equation}
    where $k_{s,\text{pen}}$ is a shear penalization constant, that we choose sufficiently large to ensure $\boldsymbol{d}\sim \boldsymbol{n}$ everywhere on the midsurface.

    \subsection{Finite-element discretization}
    At each time step, we must solve for the shell midsurface velocity $\boldsymbol{U}$ and the new director $\boldsymbol{d}$. In fact, we consider the rate of change $\dot{\boldsymbol{\beta}}$ of the $\beta$-parameters to be the main unknowns from which we also have
    \begin{equation}
        \boldsymbol{\omega}  = \begin{bmatrix}
        -\sin \beta_{01} \sin \beta_{02} &  \cos \beta_{01} \cos \beta_{02}\\
        -\cos \beta_{01} & 0 \\
        -\sin \beta_{01} \cos \beta_{02} &- \cos \beta_{01} \sin \beta_{02}
        \end{bmatrix} 
        \dot{\boldsymbol{\beta}}= \boldsymbol{W}(\boldsymbol{\beta}_0)\dot{\boldsymbol{\beta}}
    \end{equation}

    As a result, we have a system with 5 degrees of freedom to solve at each time step: three velocity components $(U_1, U_2, U_3) = \boldsymbol{U}$ and two angular rates of change $(\dot{\beta}_1,\dot{\beta}_2) = \dot{\boldsymbol{\beta}}$.
    In the present work, we do not follow the MITC discretization discussed in \cite{Hale_FEniCS-Shellls_2018} to remove shear locking but rather use a much simpler finite-element discretization, namely:
    \begin{itemize}
       \item a quadratic continuous ($P^2$-Lagrange) interpolation for the velocity field $\boldsymbol{U}$
       \item a linear interpolation with continuity enforced at the triangle mid-sides (Crouzeix-Raviart element) for the angular rates $\dot{\boldsymbol{\beta}}$
    \end{itemize}
    Such a discretization choice is extremely simple and seems free of any shear locking effects as shown in \cite{campello2003triangular} for nonlinear shells and \cite{bleyer2021novel} in the case of plate and shell limit analysis theory.

    \subsection{System evolution}
    When considering the system evolution between time $t_n$ and $t_{n+1}$, the resulting linear system for the current velocity and rate of change of director field is solved when considering the shell thickness to be fixed to its previous value i.e. $T=T_n$
    \begin{equation}
        \boldsymbol{U}_{n+1},\dot{\boldsymbol{\beta}}_{n+1} = \argmin_{\boldsymbol{U},\dot{\boldsymbol{\beta}}} \mathcal{R}( \boldsymbol{U},\dot{\boldsymbol{\beta}};T_{n})
    \label{eq:discr-Rayleighian}
    \end{equation}
    where
    \begin{equation}
        \mathcal{R}( \boldsymbol{U},\dot{\boldsymbol{\beta}};T_{n}) = \int_S \phi(\boldsymbol{U},\dot{\boldsymbol{\beta}};T_{n}) \,\dd A 
    \end{equation}
    is the total Rayleighian of the system, of surface density $\phi$ given by \eqref{eq:phi}. Here, we assume no external loading on the shell, such that all shell deformations are powered by internally generated active stresses. Note that, due to the linear expressions of the strain rates \eqref{eq:discr-Delta_ind}, \eqref{eq:discr-Omega_ind}, \eqref{eq:discr-gamma_ind} and the quadratic form of the dissipation potential $\phi$, the resulting system to solve is, in fact, linear \cite{audoly2013discrete,Turlier_BioPhys_2014}. 

    We then update the geometry in an asynchronous and Lagrangian manner, by solving the discretized thickness evolution equation and updating the midsurface position and director using \eqref{eq:director-parametrization} as follows:

    \begin{subequations}
    \begin{align}
         &\frac{T_{n+1}-T_n}{\Delta t}=-T_{n+1} a^{\alpha\beta}\Delta_{\alpha\beta,n+1}-T_{n+1} k_d + v_p (1-H_n T_{n+1} )\label{eq:thickness_discrete}
        \\
        &\boldsymbol{d}_{n+1} = \boldsymbol{d}\big(\bm{\beta}_n + \Delta t \dot{\bm{\beta}}_{n+1}\big)\label{eq:director-update}
        \\
        &\boldsymbol{x}_{n+1} = \boldsymbol{x}_{n} + \Delta t\,\boldsymbol{U}_{n+1} \label{eq:geometry-update}
    \end{align}
    \end{subequations}

     The full finite-element implementation is done within the FEniCS software package \cite{logg2012automated, alnaes2015fenics} and the meshes are produced with the Gmsh generator \cite{geuzaine2009gmsh}. In the geometry updating process, we also regularly call some remeshing procedure using the MMG platform \cite{dapogny2014three}. The Python code used for implementation in FEniCS, as well as the two following numerical illustrations may be found here \cite{githubFENICS}. Considering the symmetry of the numerical examples presented in Sec.~\ref{sec:numericalillustration}, the finite element model of the cell is modelled as one-eighth of a sphere. 

    \subsection{Volume constraint}\label{sec:VolumeConservation}

    In some numerical illustrations, such as cell division, Sec.~\ref{sec:cytokinesis}, we want to ensure that the total volume $V$ of the cell is conserved during the division. To do so, we enforce the following constraint,
\begin{equation}
\frac{d V}{d t} = \int_{\mathcal{S}} U_\alpha n_\alpha\, \dd A = 0.
\end{equation}
Enforcement of this constraint is achieved through the introduction of a single scalar Lagrange-multiplier which can be interpreted as the cell hydrostatic pressure $P$. Forming the system Lagrangian,
\begin{equation}
    \mathcal{L}(\boldsymbol{U},\dot{\boldsymbol{\beta}},P;T_{n}) = \mathcal{R}(\boldsymbol{U},\dot{\boldsymbol{\beta}};T_{n}) +  P\int_{\mathcal{S}} U_\alpha n_\alpha\, \dd A
\end{equation}
we turn the minimization problem \eqref{eq:discr-Rayleighian} into a saddle-point problem
\begin{equation}\label{eq:discr-Lagrangian}
\boldsymbol{U}_{n+1},\dot{\boldsymbol{\beta}}_{n+1},P_{n+1} = \argmax_P \argmin_{\boldsymbol{U},\dot{\boldsymbol{\beta}}} \mathcal{L}( \boldsymbol{U},\dot{\boldsymbol{\beta}},P;T_{n}).
\end{equation}

Note that the previous approach can also be easily extended to the case where the volume change rate is imposed to a given constant $\dot{V}_\text{imp}$
\begin{equation}
\frac{d V}{d t} = \int_{\mathcal{S}} U_\alpha n_\alpha\, \dd A = \dot{V}_\text{imp}
\end{equation}
by considering this new Lagrangian instead:
\begin{equation}
    \label{eq:Lagrangian-volume-change}
    \mathcal{L}(\boldsymbol{U},\dot{\boldsymbol{\beta}},P;T_{n}) = \mathcal{R}(\boldsymbol{U},\dot{\boldsymbol{\beta}};T_{n}) +  P\left(\int_{\mathcal{S}} U_\alpha n_\alpha\, \dd A - \dot{V}_\text{imp}\right).
\end{equation}


\section{Numerical illustration: osmotic shocks and cell division}
\label{sec:numericalillustration}
  
This section presents some biologically relevant model problems illustrating the partitioning of the sheet deformation between stretching and bending dominated modes. In particular, cell division and hypo-osmotic shocks are mainly dominated by tensions, whereas fast hyper-osmotic shocks will be bending-dominated. 

In the following, for the sake of simplicity we neglect local variations in the filament orientation within the cortex. We furthermore suppose that filaments remain parallel to the cell surface, which is a good approximation for the true organisation of actin in the cell cortex, as observed experimentally \cite{medalia2002macromolecular,morone2006three}. Since we neglect tangential gradients of orientation, filaments are supposed isotropically oriented along the midsurface (see Fig.~\ref{fig:nematics}). To account for the isotropic organization of filaments along the midplane, we assume a purely nematic form of the active stress tensor $Q_{ij}$ in 3D, with constant diagonal coefficients in the local midsurface basis $(\bm{a}_\alpha,\, \bm{n})$ \cite{chaikin_lubensky_1995, Salbreux_PRL_2009, Turlier_BioPhys_2014}
\begin{equation}
    Q_{ij} = \begin{pmatrix}
1/6 & 0 & 0\\
0 & 1/6 & 0 \\
0 & 0 & -1/3 
\end{pmatrix}
\end{equation}

In all generality $Q_{ij}$ may also have an isotropic contribution. However such isotropic term would appear only in active bending moments, but not in membrane resultants. We will see that in our two examples below (osmotic shock and cell division) that active bending moments play a negligible role in the shape dynamics, and we therefore ignored such contribution.
To explicitly consider filament reorganization one may have to write down and expand a dynamic equation for the spatiotemporal evolution of this nematic tensor and its coupling to velocity gradients \cite{Kruse_EPJE_2005,Salbreux_PRL_2009,Prost_Nature_2015,napoli2016hydrodynamic,metselaar2019topology}, which is out of the scope of this manuscript.

\subsection{Osmotic shock}
\begin{figure}[ht]
\centering
\includegraphics[scale=1]{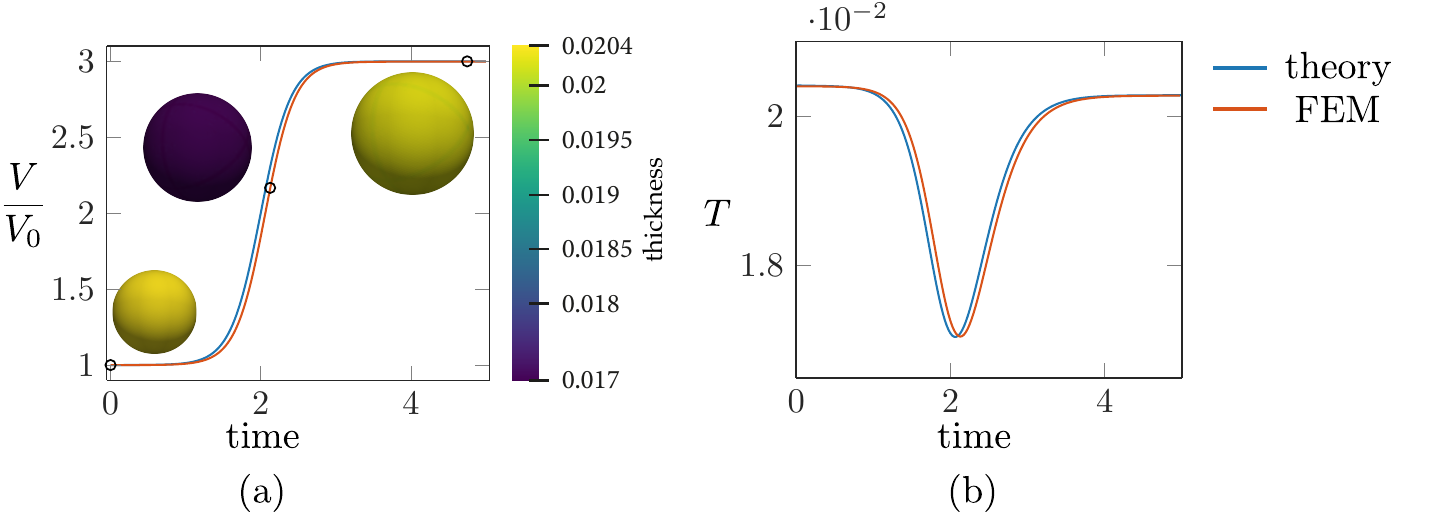}
\caption{ Hyposmotic shock. (a) Imposed volume, inserts represent the cell in scaled size and the colors represent the cell's thickness. (b) The time evolution of the thickness. Numerical values are summarized in table~\ref{table}. }\label{fig:OsmoticShock_Inflating}
\end{figure}

\begin{figure}[ht]
\centering
\includegraphics[width=\textwidth]{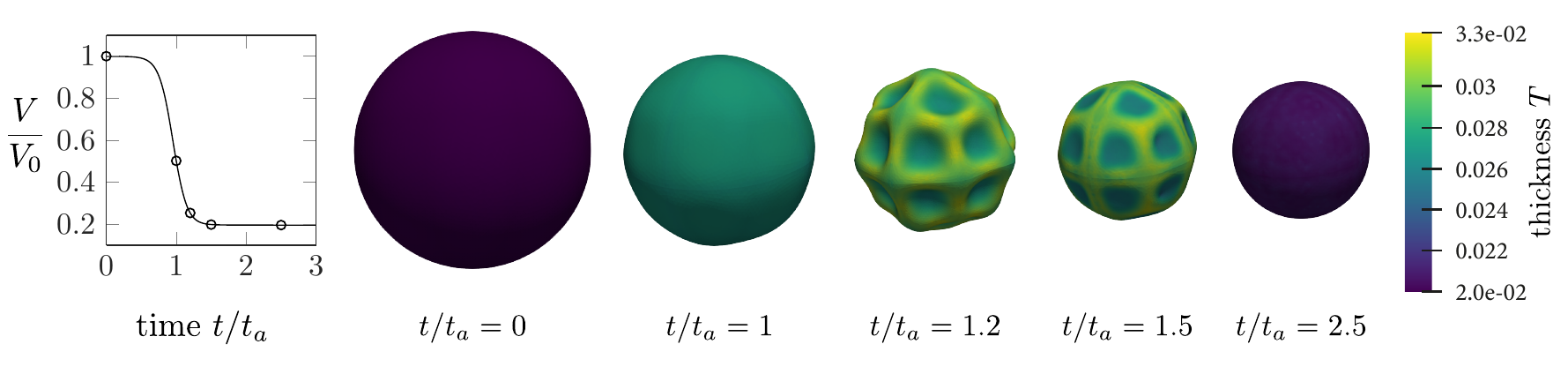}
\caption{Shape dynamics timecourse of a cell subjected to a hyperosmotic shock. In the left is the \textit{imposed} volume variation over time. The snapshots correspond to the cell shape at successive times. The color-scheme represents the local thickness of the cell. $t_a \equiv \mu / \zeta$ is the typical deformation timescale. }\label{fig:OsmoticShock}
\end{figure}

Here, we mimic the response of a suspended cell to rapid osmotic shocks. Considering a cell in osmotic equilibrium with an external medium, a sudden change in solute concentration creates an osmotic imbalance that will generate a rapid flux of water across the cell membrane, changing its volume until a novel equilibration of concentrations is reached. If the solute concentration in the medium is \textit{increased}, the cell is subjected to a \textit{hyperosmotic shock}, causing an outward flux of water and cell shrinkage. In contrast, a \textit{decrease} in solute concentration leads to an inward flux in the cell called \textit{hypo-osmotic shock}, leading to its expansion. For the sake of simplicity, we don't intend here to solve for a realistic pump-leak model of cell volume control, involving electroosmotic effects \cite{Hoppensteadt2002,mori2012mathematical,kay2017cells}, but we impose instead directly the variation of a target volume $V(t)$ using the constraint \eqref{eq:Lagrangian-volume-change} (see Figs.~\ref{fig:OsmoticShock_Inflating}a and \ref{fig:OsmoticShock}). 

We first consider an hypo-osmotic shock, where a solution may be calculated analytically. We suppose that the cell's response is mainly in the membrane regime, and that the volume can be externally controlled and is given by $V(t)$. Because the deformation is spherically symmetric, the time evolution for the radius of the expanding cell reduces to $R(t) = \sqrt[3]{\frac{3 V(t)}{4 \pi}}$. In spherical coordinates, the metric and curvature tensors are $a_{\alpha\beta} = \text{diag}(R^2,\, R^2\sin^2 \theta)$, and $b_{\alpha\beta} = \text{diag}(-R, \, -R\sin^2 \theta) $, respectively. Moreover, since $\Delta_{\alpha\beta} = \frac{1}{2} \frac{d a_{\alpha\beta}}{dt}$, we have for isotropic deformations $\Delta_{\alpha\beta} = \text{diag}(R \dot{R},\, R \dot{R} \sin^2 \theta)$.



The thickness evolution, Eq.~\eqref{eq:thickness}, can be readily written as
\begin{equation}\label{eq:ode_T}
   \frac{\dot{T}}{T}  + 2 \frac{\dot{R}}{R} +  \left(k_d-\frac{v_p}{T}\right)  - \frac{v_p}{R} =0.
\end{equation}
The first-order differential equation \eqref{eq:ode_T}, subjected to the initial condition $T(t \!=\! 0) = T_0$ can be analytically solved for a given $V(t)$, consequently, given $R(t)$. The steady-state thickness at a given cell radius $R$ is given by $T_0 = \frac{R v_p}{k_d R - v_p}$. It slightly differs from the value $v_p/k_d$ value defined in \cite{Turlier_BioPhys_2014, Salbreux_PRL_2009}, which is valid only in the limit of a flat plate $R\rightarrow \infty$. It is corrected here by a curvature term $ -v_p H T$ in the thickness evolution, that comes from the asymmetry of the polymerization and the curved geometry of the sheet.

Such simplified system can be used to validate the FEM approach: we impose a sigmoid volume variation $V(t)$ and compare the analytical solution of \eqref{eq:ode_T} for the thickness $T(t)$  with the solution obtained through finite elements under the same set of parameters, see Fig.~\ref{fig:OsmoticShock_Inflating}. 


\paragraph{Young-Laplace equation for thin shells}
In a spherical symmetric system, assuming that tangential tension gradients and bending moments may be neglected, an analog of the classic Young-Laplace equation can be retrieved for thin shells from the out-of-plane equation \eqref{eq:eq_out_plan}, 
\begin{equation}
    \label{eq:Laplace-equation}
     \left(1+\frac{T}{R}\right)P = \frac{2N}{R} = \frac{2}{R}\left[4 \mu T \frac{\dot{R}}{R}+ 2 \mu T\left(k_d + \frac{v_p}{T}\right) + \frac{1}{2} \zeta T \right],
\end{equation}
where $P$ is the internal pressure that maintains the cell volume constant (external pressure is supposed zero), and $N$ the tension supposed isotropic and homogeneous along the membrane. It is interesting to note the geometric correction to Laplace's law in $T/R = \mathcal{O}(\e)$ in front of the pressure, that is usually neglected for very thin shells or membranes.

In the approximation of slow volume change ($\dot{T}=0$, $\dot{R} = 0$) the above coupled equations at steady-state reduce to
\begin{equation}
\label{eq:stationary-Laplace-law}
    P = \frac{2}{R}T_{0}^{\rm{flat}}\left(\frac{1}{2}\zeta + 4\mu k_d\frac{T_0^{\rm{flat}}}{R} \right),
\end{equation}
where $T_0^{\rm{flat}} \equiv \frac{v_p}{k_d}$ is the stationary thickness in the limit of a flat shell $R\rightarrow \infty$ with no flow. One can remark that the typical Laplace's formula used in spherical geometry $P = \frac{2}{R}\frac{T_0^{\rm{flat}}\zeta}{2}$ \cite{Salbreux_PRL_2009,Turlier_BioPhys_2014} is corrected by a term of order $T/R = \mathcal{O}(\e)$ proportional to the turnover rate $k_d$.

\paragraph{Hyperosmotic shock} To probe our model in situations where bending dissipation may become non-negligible, we study the effects of a rapid hyperosmotic shock on a cell. Here, turnover acts as a stabilizing agent: if the volume variation is slower than the typical turnover timescale, the cell cortex will have the time to rearrange and adapt the cell's shape to its new volume without buckling. We thus apply a sudden drop in volume that we model by a sigmoid function, as illustrated in Fig.~\ref{fig:OsmoticShock}. As the cortex does not have time to depolymerize homogeneously, folding induces spatial variation in the thickness correlated to local variations in curvature.
    
\begin{figure}[h!]
\centering
\includegraphics[scale=1]{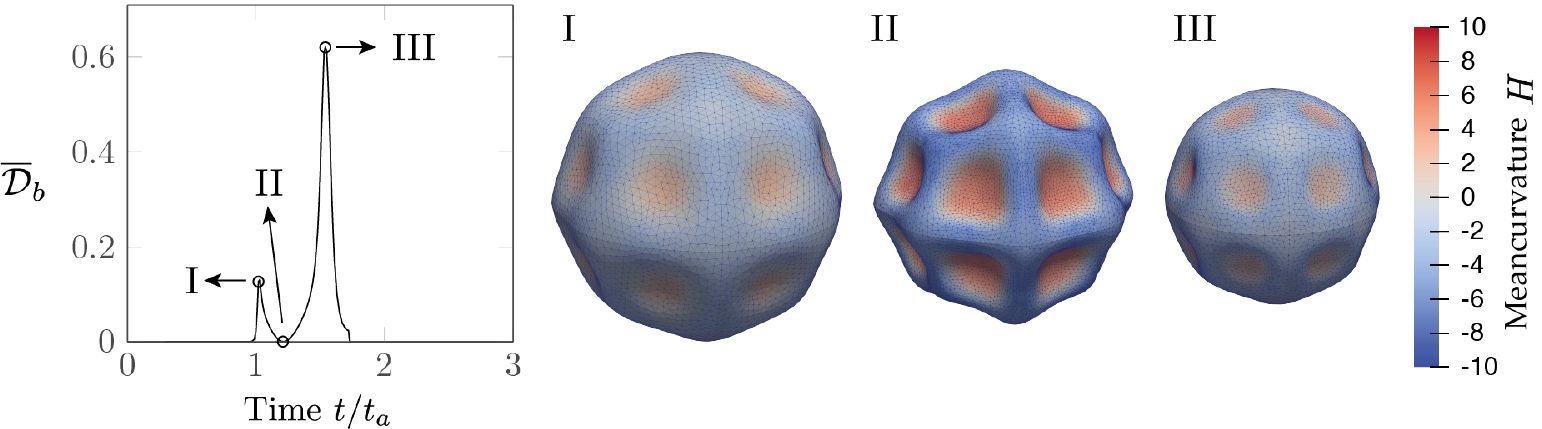}
\caption{Plot of the relative contribution of the bending dissipation to the total dissipation. The two peaks (I and III) represent the instants in time for which the rate of curvature change is the highest. Even though at intermediary time II the cell shape is overall much more curved, bending dissipation is the lowest. }\label{fig:OsmoticShock_D}
\end{figure}

    In our simulations shown in Fig.~\ref{fig:OsmoticShock}, we find numerically that active bending terms play a negligible role, contributing to less than $0.1\%$ of the total power input.
    However, passive bending terms are not negligible.
    To compare the relative contributions to the dissipation of bending and stretching rates, we define the following non-dimensional parameter
    \begin{equation}
        \overline{\mathcal{D}}_b = \frac{\mathcal{D}_{b}}{\mathcal{D}_{s}+\mathcal{D}_{b}} = \frac{\iint_{\mathcal{S}} \mu \mathcal{A}^{\alpha\beta\gamma\delta}  \frac{1}{3} T^3 \Omega_{\alpha\beta} \Omega_{\gamma\delta} \dd A}{\iint_{\mathcal{S}} \left[\mu \mathcal{A}^{\alpha\beta\gamma\delta} (4 T \Delta_{\alpha\beta} \Delta_{\gamma\delta} + \frac{1}{3} T^3 \Omega_{\alpha\beta} \Omega_{\gamma\delta}) \right]\dd A},
    \end{equation}
    which measures the relative weight of bending dissipation $\mathcal{D}_b = \iint_{\mathcal{S}} \mu \mathcal{A}^{\alpha\beta\gamma\delta}  \frac{1}{3} T^3 \Omega_{\alpha\beta} \Omega_{\gamma\delta} \dd A$ to the total viscous dissipation in the shell $\mathcal{D}= \mathcal{D}_{s}+\mathcal{D}_{b}$, where $\mathcal{D}_s = \iint_{\mathcal{S}} 4 \mu T \mathcal{A}^{\alpha\beta\gamma\delta} \Delta_{\alpha\beta} \Delta_{\gamma\delta} \dd A$ is the dissipation due to stretching. The time-course of the parameter $\overline{\mathcal{D}}_b$ is plotted in Fig.~\ref{fig:OsmoticShock_D}. Two distinctive peaks indicate a significant contribution of bending to dissipation, which may reach up to $60\%$ of the total. Interestingly, these two peaks, indicated by (I) and (III), do not correspond to the most folded cell shape, which rather lies at the minimum of the curve between these two maxima, as indicated by (II). This is because, for a thin viscous sheet, dissipation increases with the rate of curvature change, but not with the curvature itself, like one may be used to in elasticity. Therefore, maintaining highly curved cell shapes does not cost energy, but dissipation shall oppose rapid shape changes. 

\subsection{Cell division}
\label{sec:cytokinesis}
Cell division encompasses two main processes in biology: mitosis, which corresponds to the even splitting of genetic material between the two future daughter cells, and cytokinesis, whereby the mother cell cytoplasm is physically cleaved in two \cite{green2012cytokinesis}. In the following, we focus on the cell's shape deformation during cytokinesis. In animal cells, cytokinesis is achieved through the ingression of a circumferential actomyosin furrow located at the equator of the cell, while cortical tension at cell poles is resisting the deformation imposed by increasing cell pressure \cite{sedzinski2011polar, Turlier_BioPhys_2014}. Various mechanical models of cell deformation over cytokinesis have been proposed in the last decades, with different assumptions about the mechanical nature of the cortical material: contractile \cite{greenspan1978fluid}, elastic \cite{pujara:1979}, viscous \cite{zinemanas1988viscous} or viscoelastic \cite{akkacs1980biomechanics}. Other approaches have considered the lipid bilayer mechanics only and attributed most of the mechanical dissipation to cytoplasmic and surrounding medium fluids \cite{he1997mechanics, Li:2016dw}. In some of these models, a contractile ring structure of different nature is supposed to provide the constricting force at the cell equator \cite{pujara:1979, akkacs1980biomechanics}, while in others, the furrowing is the natural result of higher but continuous activity in the equatorial region, that may trigger possible structural rearrangements of the continuous cortical surface \cite{greenspan1978fluid, zinemanas1988viscous}. In agreement with early hypotheses \cite{White:1983th}, recent active-gel models of cytokinesis \cite{Turlier_BioPhys_2014} and experiments \cite{sedzinski2011polar, Spira:2017} all point towards the continuum character of the cortex, that may flow in a Marangoni-like process toward the equator as a result of contractile gradients and can reorganize structurally along the cell's surface in response to extensile and compressive forces. In fact, the hypothesis of surface tension gradients driving division is ancient and goes back to early observations of dividing marine eggs \cite{butschli:1876, McClendon:1913gu,spek:1918} and to the similarity of their shape to one of the olive-oil droplets in response to surfactants \cite{greenspan1978fluid}. A gradient of contractility centered at the equator has indeed been measured from the cell poles to the cell equator in the form of a Gaussian-like spatial distribution of active RhoA, a protein upstream of actomyosin contractility \cite{bement2005rhoA}.

\begin{figure}[h!]
\centering
\includegraphics[width=0.98\textwidth]{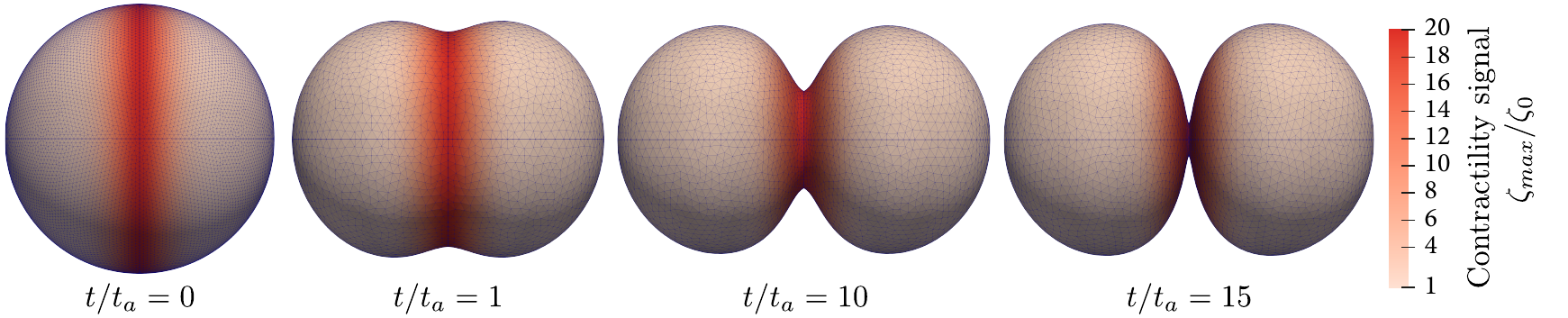}
\caption{Dynamic shape changes of a cell during cell division (cytokinesis) illustrated by successive timepoints of furrow constriction. The color-scheme refers to the rescaled contractile activity signal $\zeta$ \eqref{eq:cytokinesis_activity_signal}.} \label{fig:Cytokinesis}
\end{figure}

\begin{figure}[h!]
\centering
\includegraphics[width=0.98\textwidth]{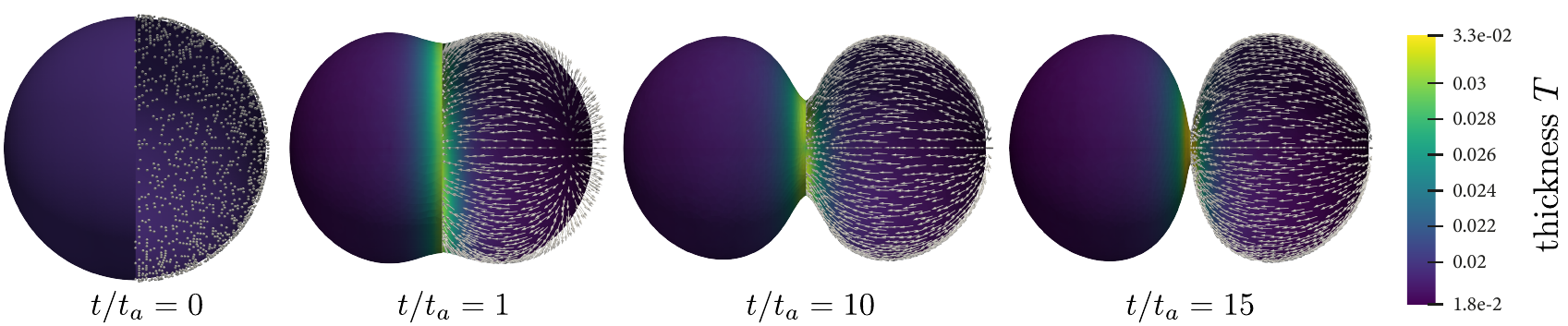}
\caption{Dynamic evolution of the thickness and velocity field during cell division (cytokinesis) illustrated by successive timepoints of furrow constriction. Arrows represent the direction of the velocity field $\vec{u}$. The color-scheme refers to the cortex thickness.}\label{fig:CytokinesisThickness}
\end{figure}

Following \cite{Turlier_BioPhys_2014}, we start from a spherical cell of radius $R_0$ and mimic the RhoA gradient by imposing a Gaussian distribution of active stress along a fixed ambient coordinate direction $x_\alpha$ centered at the equator,
\begin{equation}
\label{eq:cytokinesis_activity_signal}
   \zeta = \zeta_0 + (\delta\zeta-\zeta_0) e^{-\frac{1}{2}\frac{x_\alpha^2}{w^2}}
\end{equation}
where $\zeta_0$ is the basal level of stress, that creates an active tension at poles, and $\delta\zeta$ is called equatorial \textit{overactivity} that expands over a typical width $w$. In the initial state, the cell is a sphere of $R_0$ and uniform thickness $T = v_p/k_d$, which is subjected to an uniform contractile activity $\zeta_0$. A typical deformation timescale in this system is the ratio of the viscosity to the contractile stress $t_a \equiv \mu / \zeta$. Typical numerical values for all these parameters are extracted from experimental measurements referenced in the literature and may be found in Table \ref{table}. Following the rise of a Gaussian over-activity, the equatorial region of the cell becomes more contractile and generates cortical flows toward the equator. This triggers a local increase in the thickness, which is partially counterbalanced by depolymerization, as found in the mass balance Eq.~\eqref{eq:mass_conserv}. Material turnover is indeed the essential factor ensuring the overall stability of the thickness evolution in presence of contractility or thickness gradients. The accumulation of matter in the furrow, here represented by the increase in thickness, increases further the contractile stress around the equator\footnote{Simple stability analysis of a flat shell under uniform active stress $\zeta$ shows that the thickness is stable if $\zeta \leq 4 \mu k_d$. If the additional tension generated by turnover $2Ta^{\alpha\beta}(kd-v_p/T)$ is neglected, the threshold for stability doubles to $\zeta \leq 8 \mu k_d$, which shows that the turnover stress has a destabilizing effect.}. The resulting \textit{contractile ring} then pinches the cell, leading to its constriction. The time course for such dynamics is shown on Fig.~\ref{fig:Cytokinesis}. The spatial distribution of contractility is assumed here to be time-independent, but it may slightly vary with cell shape changes \cite{bement2005rhoA}. The spatiotemporal evolution of the thickness and of the midsurface velocity field are illustrated at corresponding timepoints in Fig.~\ref{fig:CytokinesisThickness}.

\begin{figure}[h!]
    \begin{subfigure}[h]{0.45\textwidth}
    \includegraphics[scale=1]{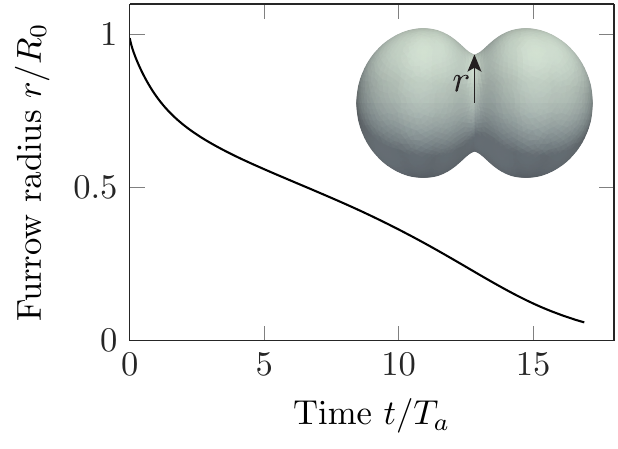}
    \caption{}\label{fig:FurrowRadius}
    \end{subfigure}
    \begin{subfigure}[h]{0.49\textwidth}
    \includegraphics[scale=1]{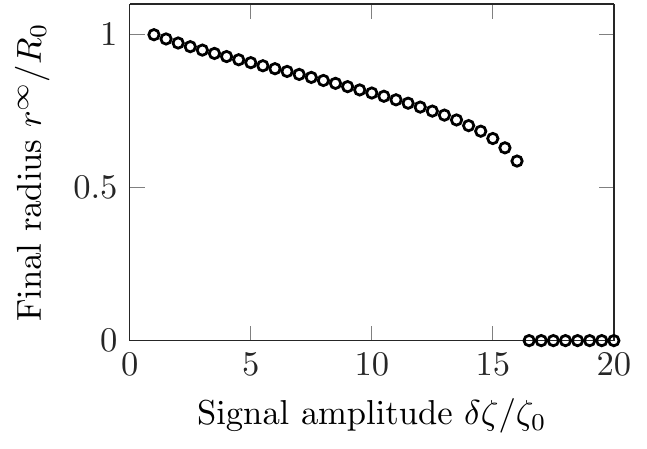}\caption{}\label{fig:Furrow_threshold}
    \end{subfigure}
    \caption{Furrow constriction dynamics. (a) Time evolution of the furrow radius. (b) Bifurcation diagram representing the final furrow radius as a function of the amplitude of the equatorial overactivity.}
\end{figure}

    As for the osmotic shocks, we find numerically that active bending plays a fully negligible role. But, in contrast to the hypoosmotic shock, bending dissipation remains here also largely negligible during the time-course of cell division (not shown), and the cell adopts very curved shapes in tje final stages of cytokinesis with almost no associated dissipation. Cytokinesis may therefore be well described by a simple membrane theory, where tension is the dominating phenomena, which justifies \textit{a posteriori} the approach of \cite{Turlier_BioPhys_2014}, where an axisymmetric active viscous membrane model was employed to simulate cytokinesis. As originally predicted in \cite{Turlier_BioPhys_2014}, we show that full constriction of the furrow, leading to the completion of cell division, depends on the level of equatorial over-activity $ \zeta/\zeta_0$, presenting a threshold $\zeta_c$, below which the full constriction fails. We plot in Fig.~\ref{fig:Furrow_threshold}, the bifurcation diagram for physiological values of the parameters. Each point in the figure represents the final radius following the dynamics as shown in Fig.~\ref{fig:FurrowRadius}, with variable signal over-activity.

\begin{figure}[h!]
\centering
\includegraphics[scale=1]{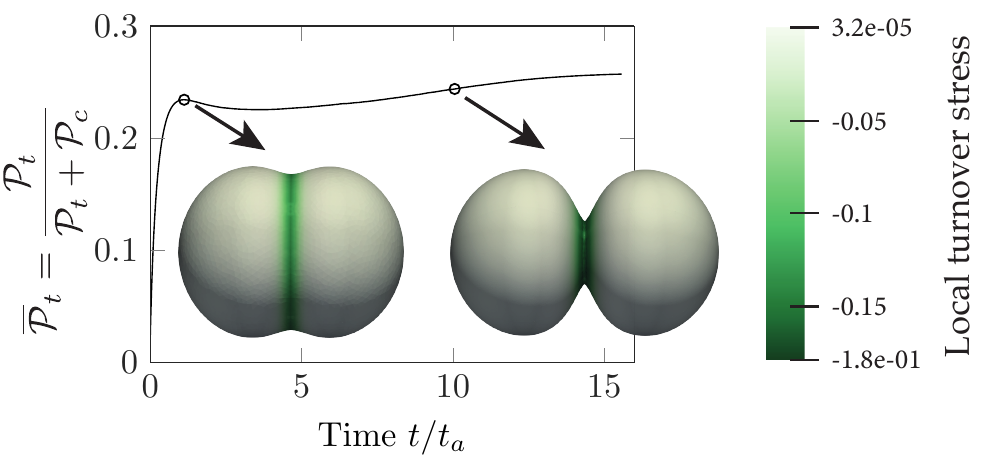}
\caption{Relative contribution of the (integrated) power input of turnover with respect to the total power input due to both turnover and contractility as function of time. The insets show that most of the stress generated by the polymerization is localized at the furrow.}
\label{fig:Turnover}
\end{figure}

    Even though the obtained results in \cite{Turlier_BioPhys_2014} were qualitatively correct, capturing well the shape changes for axisymmetric cell division, the model neglected an essential stress contribution, generated by active material (de)-polymerization. To compare the relative contribution of the turnover to the total input power, we define the following dimensionless parameter
    \begin{equation}
       \overline{\mathcal{P}}_{t} = \frac{\mathcal{P}_{t}}{\mathcal{P}_{t}+\mathcal{P}_{c}} = \frac{ \iint_{\mathcal{S}}\mu T a^{\alpha\beta} \left( k_d -\frac{v_p}{T}\right)\Delta_{\alpha\beta} \dd A}{ \iint_{\mathcal{S}}\left[\mu T a^{\alpha\beta} \left( k_d -\frac{v_p}{T}\right)\Delta_{\alpha\beta} + \frac{1}{2} T \mathcal{Q}^{\alpha\beta} \zeta \Delta_{\alpha\beta}\right] \dd A},
    \end{equation}
   where $\mathcal{P}_{t} = \iint_{\mathcal{S}} \mu T a^{\alpha\beta} \left( k_d -\frac{v_p}{T}\right)\Delta_{\alpha\beta} \dd A$, and $\mathcal{P}_{c} = \iint_{\mathcal{S}} \frac{1}{2} T \mathcal{Q}^{\alpha\beta} \zeta \Delta_{\alpha\beta} \dd A$ are active power contributions due to turnover and contractility, respectively. Note that we did not include the active bending term in $\overline{\mathcal{P}}_{t}$, as its contribution to the active power input is negligible. The stress generated by the polymerization comes from the local difference in thickness $k_d\left(1-\frac{T_0}{T}\right)$ created by the cortical flows with a reference thickness $T_0=\frac{v_p}{k_d}$ for a flat plate with vanishing flow. Since the variation in thickness is highly concentrated around the equator of the cell, most of the active power input due to the turnover stress will be concentrated in this region and accounts to up to 25$\%$ of total (integrated) active power input, as shown in  Fig.~\ref{fig:Turnover}.
        

\section{Discussion}
\label{sec:conclusion}

The lack of generic and flexible theoretical and numerical frameworks to realistically simulate the 3D shape changes of single or multiple interacting cells has been limiting our ability to characterize complex morphogenetic events, such as asymmetric cell divisions \cite{Henry1986, maddox2007anillin, Ou2010, herszterg2013interplay, Roubinet:2017ce} and rapid multicellular reorganizations \cite{singh2014coupling, Jelier:2016kk, Sugioka2018}, which are fundamental to early embryogenesis and tissue morphogenesis. To fill this gap, we derived in this paper a viscous active thin shell theory for the cell cortex in general curvilinear coordinates, and we proposed a numerical implementation with the finite-element library FEniCS \cite{logg2012automated, alnaes2015fenics}, which we hope will facilitate the spread, reuse, modification, and refinement of our model.

We have used systematic asymptotic expansions to derive leading-order constitutive equations governing the dynamics of the actomyosin cortex. The main geometrical assumption is the slenderness of the cortex, meaning that the characteristic thickness $T_0$ must be small relative to a characteristic size of the cell, typically its curvature radius. Our main physical assumption is to describe the actomyosin cortex as a continuous medium, relying on hydrodynamics active-gel models\cite{Kruse_EPJE_2005, Prost_Nature_2015}. Here, we chose to model the actomyosin material as a single incompressible active fluid phase following previous approaches \cite{Salbreux_PRL_2009, Turlier_BioPhys_2014}. However, the full rheological and dynamic complexity of actomyosin networks may not be fully encompassed by such a simple constitutive equation. The asymptotic expansion approach may provide a generic and rigorous method to derive alternative thin shell models based on refined descriptions of actomyosin rheology. A first extension would be to account explicitly for the dynamics of the nematic order of filaments. Active nematic gel theories \cite{Prost_Nature_2015,Salbreux_PRL_2009} provide a convenient framework to describe the bidirectional coupling between nematic order and surface flows. Our derivation was done with a generic active tensor $Q_{\alpha\beta}$. It was then supposed nematic in the two examples to account for the approximate alignment of filaments parallel to the midsurface, while we neglected dynamic filament reorganization in this tangent plane. To obtain a proper nematic active thin viscous shell theory, one should write explicitly a bulk dynamic equation for $Q_{\alpha\beta}$ and reduce it into an effective 2-dimensional equation defined on the midplane. Our theory may also be readily extended to account for possible transverse gradients in myosin \cite{truong2021extent}, that could generate additional active torques. We assumed that actomyosin is incompressible, but this material may be better described as a permeating gel by introducing a local actin density \cite{CallanJones:2011iz, joanny2013actin}. Describing the cortex as a gel seems important to better assess the contribution of turnover to stress resultants, which relies here importantly on the incompressibility condition. By essence, our model was designed as purely viscous and is well-suited to describe shapes changes on timescales much larger than a typical actomyosin turnover time $k_d^{-1}$. But it could also be extended to describe the viscoelastic response of the cortex on shorter timescales \cite{Khalilgharibi:2019jt}, although this may entail the more difficult handling of power-law rheological models \cite{bonfanti2020fractional}. All these possible extensions rely generally on bulk descriptions of actomyosin mechanics that do not derive directly from microscopic descriptions of the network. Ample numerical and theoretical effort will also be required in the field to bridge these scales quantitatively.

Our thin shell theory accounts for viscous stretching and bending resistance, making it relevant to describe rapid curvature changes. Although viscous in nature, the Stokes-Rayleigh analogy allows for interpreting our model in terms of the classic theory of elastic shells, and the equations we obtained may be classified into Koiter-like shell models.
In contrast to alternative direct active surface approaches \cite{salbreux2017mechanics}, the derivation of a proper thin shell theory starting from three-dimensional constitutive equations, shows its relevance for the cortex through two important predictions:
\begin{itemize}
\itemsep0em 
    \item First, the balance between polymerization and depolymerization is predicted to create contractile or extensile tensions and may also oppose or promote the bending of the layer. This effect depends inherently on the shell character of the cortex and relies primarily on our hypothesis of an incompressible phase and on the fact that the cortex grows from the plasma membrane, creating an inward addition of material that may translate into longitudinal viscous flows, and hence into turnover-generated viscous stresses. It follows that active cell deformations depend not only on the contractile activity of myosin motors but also on the speed of actin (de)polymerization, which is particularly important to relate cell shape changes to its molecular regulation. Such contribution has been neglected in previous theories of active membranes \cite{Salbreux_PRL_2009, Turlier_BioPhys_2014, Arroyo_JFM_2019}.  Yet, using experimentally measured parameters, we find that up to $25\%$ of the active power input may come from turnover during cytokinesis and that it largely contributes to the equatorial contractile force that divides the cell.
    \item Second, a proper thin shell model with bending resistance is necessary to account for rapid changes in cell curvatures, as one may observe in hyperosmotic shocks. Here, the asymptotic approach allowed us to calculate prefactors for active bending terms \eqref{eq:M_composite}, which would have been otherwise difficult to write down directly. Yet, in the two numerical examples we chose, we found that active bending was largely negligible.  
\end{itemize} 

From a numerical perspective, we provided an easily modifiable Python implementation of the model within the general-purpose finite-element library FEniCS. We proposed a numerical scheme based on a penalization method that circumvents the need for $C^1$ elements to use more simple $C^0$ elements. In the cortex, gradients in contractile stress can lead to strong tangential motions, and convergent flows, which create excessive distortions of the computational mesh linked to the material \cite{Turlier_BioPhys_2014} (see Fig.~\ref{fig:Cytokinesis}). In the present work, such significant distortions were circumvented through repeated remeshing of the computational mesh and interpolation of the fields. This solution may prove costly for larger meshes, and a more appropriate alternative using an Arbitrary Lagrangian-Eulerian (ALE) representation \cite{donea2004arbitrary} may be used, in which one is free to decouple the movement of the Lagrangian particles from the movement of the mesh. Such an approach has been previously employed for the simulation of fluid membranes \cite{Arroyo_JFM_2019, Sahu_JCP_2020} and our description can be very easily translated into a similar ALE formulation. 

Our finite-element numerical scheme may be adapted in the future to tackle more complex morphogenetic processes. In particular, one can readily envision the study of the cell division when the cell is confined to a rigid shell, as happens for instance in sea-urchin or worm eggs, such as \textit{C elegans}, where such confinement leads to two tightly apposed membranes in the furrowing region \cite{carvalho2009structural}. During division and migration, cortical flows have been long predicted to reorient actin filaments \cite{White:1983th, reymann2016cortical,Spira:2017}, which in turn may influence cortical contractility \cite{greenspan1978fluid,zinemanas1988viscous}. Interestingly, Rayleigh variational formulations for 2D nematic surfaces are already available \cite{napoli2016hydrodynamic} but solving numerically for the coupled viscous-nematic dynamics remains a challenging task \cite{metselaar2019topology}, that may require specific finite-element approximations to implement tensorial dynamics on moving boundaries \cite{nestler2019finite, torres2020approximation}. Another possible extension of our numerical approach is the coupling of cortical mechanics with regulatory proteins lying at the plasma membrane, which may diffuse, react and be advected by cortical flows to generate complex morphogenetic patterns \cite{carroll2003exploring, bement2015activator, Gross:2017fg, bischof2017cdk1, Mietke_PNAS_2019, Yin:2021gw}. Finally, the generalization our model to multicellular systems, by solving, for instance, similar equations on non-manifold triangular meshes \cite{Maitre2016}, would allow us to envision very realistic simulations of cell mechanics in early embryo development.

 \section*{Acknowledgements}
 This project was supported by the Bettencourt-Schueller Foundation and the CNRS/Inserm ATIP-Avenir program. HBR received financial support from the Labex MemoLife under the program "Investissements d'Avenir" ANR-10-LABX-54. We thank N. Ribe and C. Maurini for helpful early discussions.

\bibliographystyle{elsarticle-num}

\newpage
\bibliography{bibfile}

\pagebreak
\appendix
\label{sec:Appendix}
{\Large \textbf{Appendix}}

\section{Asymptotic expansion: Inextensional limit}\label{app:Asymptotic_bending}
The appropriate scaling for the inextensional regime \eqref{eq:scaling_bending} gives the following expansion
\begin{equation}\label{eq:asymptoticbending}
u_i=\frac{PL}{\mu \e^3}\sum_{m,n=0}\varepsilon^m\hat{z}^n u_i^{(mn)},
\quad
p=\frac{P}{\e^2}\sum_{m,n=0}\varepsilon^m\hat{z}^n p^{(mn)}.
\end{equation}

It will be useful to introduce the dimensionless strain rate tensors, and rate of curvature change 
$$\Delta^{(m0)}_{\alpha\beta}=\frac{1}{2}(u_{\alpha\vert\beta}^{(m0)}+u_{\beta\vert\alpha}^{(m0)})-\hb_{\alpha\beta} u_3^{(m0)},$$ 
$$\Omega_{\alpha\beta}^{(m0)}=u^{(m0)}_{3,\alpha\vert\beta}+b^\lambda_{\beta} u_{\lambda\vert\alpha}^{(m0)}+b^\lambda_{\beta\vert\alpha} u_\lambda^{(m0)}+\hat{b}^\gamma_\beta u_{\gamma\vert\alpha}^{(m0)}-\hat{b}^\gamma_\beta b_{\alpha\gamma} u^{(m0)}_3.$$

The strain rate tensor is obtained similarly by inserting \eqref{eq:asymptoticbending} into \eqref{eq:strainrate}
\begin{equation}\label{eq:strainrateexpansionbending}
\begin{split}
2e_{\alpha \beta} & = \frac{P}{\mu \e^3}\sum_{m,n=0}\varepsilon^m\hat{z}^n
\left[ u_{\alpha\vert\beta}^{(mn)}+u_{\beta\vert\alpha}^{(mn)}-2\hb_{\alpha\beta}u_3^{(mn)} \right.
 +\left.\e\hz(2\hb^\lambda_\beta\hb_{\lambda\alpha}u_3^{(mn)}-\hb^{\lambda}_\alpha u_{\lambda\vert\beta}^{(mn)}-\hb^{\lambda}_\beta u_{\lambda\vert\alpha}^{(mn)})\right],
\end{split}
\end{equation}
which we then use to obtain the non-symmetric stress tensor 
\begin{equation}
\begin{split}
\frac{\sigma^{\alpha \beta}}{P}=& \frac{1}{\e^3}\sum_{m,n=0}\varepsilon^m\hat{z}^n\left[ \phi^{(mn)}_{0}+\e\hz\phi^{(mn)}_{1}+ \e^2\hz^2\phi^{(mn)}_{2}+\mathcal{O}(\e^3)\right] + 
\\
& \e^{-2} a^{\alpha \mu}a^{\beta \delta}\hat \zeta Q_{\mu \delta}+
\\
&  \e^{-1}\hat z \,(a^{\alpha\mu}\hb^{\beta\delta}+2a^{\beta\delta} \hb^{\alpha\mu}-2\hat{H}a^{\alpha\mu}a^{\beta\delta})\hat \zeta Q_{\mu\delta}+
\\
& \e^{0} \hat z^2 \,( \hat{G} a^{\alpha\mu}a^{\beta\delta}-2 \hat{H} a^{\alpha\mu}b^{\beta\delta} - 4 \hat{H} b^{\alpha\mu}a^{\beta\delta} + b^{\alpha\mu}b^{\beta\delta} -2 a^{\alpha\mu}\hb^{\delta\gamma} \hb^\beta_\gamma) \hat \zeta Q_{\mu\delta},  
\end{split}
\end{equation}
where
\begin{equation}
\phi^{(mn)}_{0}=-a^{\alpha\beta} p^{(mn)}\e+2 a^{\alpha\mu}a^{\beta\delta}\left[\frac{1}{2}(u_{\mu\vert\delta}^{(mn)}+u_{\delta\vert\mu}^{(mn)})-\hb_{\mu\delta}u_3^{(mn)}\right],
\end{equation}
\begin{equation}
\begin{split}
\phi^{(mn)}_{1}= -(\hb^{\alpha\beta}-2\hat{H}a^{\alpha\beta}) p^{(mn)}\e
+4\, (a^{\beta\delta} \hb^{\alpha\mu} - \hat{H}a^{\alpha\mu}a^{\beta\delta})\left[\frac{1}{2}(u_{\mu\vert\delta}^{(mn)}+u_{\delta\vert\mu}^{(mn)})-\hb_{\mu\delta}u_3^{(mn)}\right],
\end{split}
\end{equation}
and
\begin{equation}
\begin{split}
\phi^{(mn)}_{2}=& (-\hat{G} a^{\alpha\beta }+2 \hat{H}\hb^{\alpha\beta}+2\hb^{\alpha\gamma} \hb_{\gamma}^\beta) p^{(mn)}\e
+
\\
& (2 \hat{G} a^{\alpha\mu}a^{\beta\delta}+4 b^{\alpha\mu}b^{\beta\delta} - 4H a^{\alpha\mu}b^{\beta\delta}-8a^{\alpha\mu}\hb^{\delta\gamma} \hb^\beta_\gamma -8 a^{\beta\delta}\hb^{\mu\gamma} \hb^\alpha_\gamma)\left[\frac{1}{2}(u_{\mu\vert\delta}^{(mn)}+u_{\delta\vert\mu}^{(mn)})-\hb_{\mu\delta}u_3^{(mn)}\right].
\end{split}
\end{equation}
Notice the extra $\e$ in front of each $p^{(mn)}$  term, which comes from the appropriate scaling for the inextensional limit. The remaining components for the stress tensor are
\begin{equation}
\begin{split}
\sigma^{\alpha3}=\frac{P}{\e^3}\sum_{m,n=0}\varepsilon^m\hat{z}^n  a^{\alpha\gamma}& [u_{\gamma}^{(m+1n+1)}(n+1)+u_{3,\gamma}^{(mn)}+\hat{b}^{\lambda}_\gamma u_\lambda^{(mn)} 
\\
&+ \e \hat{z}[-(n+1)\hb^{\alpha \gamma} u_\gamma^{(m+1 \, n+1)} + 2(\hb^{\alpha\gamma}-\hat{H}a^{\alpha\gamma})(u_\gamma^{(m+1 \, n+1)}+u_{3\vert \gamma}^{(mn)}) ]+\mathcal{O}(\e^2)] , 
\end{split}
\end{equation}
and
\begin{equation}
\begin{split}
\sigma^{33}=\frac{P}{\e^3}\sum_{m,n=0} \varepsilon^m\hat{z}^n 
\left[-p^{(mn)}\e+  2 (n+1)u_3^{(m+1n+1)}\right] (1-2\hat{H}\hz\e+\hat{G}\hz^2\e^2)   +\e^{-2}\hat \zeta Q_{33}(1-2\hat{H}\hz\e+\hat{G}\hz^2\e^2)
\end{split}
\end{equation}

The stress resultant, integrated tension along the thickness is
\begin{equation*}
\begin{split}
\frac{n^{\alpha\beta}}{PL}
=&\frac{1}{\e^2}  \sum_{m,n=0}\varepsilon^m \left[ \frac{2}{2n+1} \left(\frac{\hat{T}}{2}\right)^{2n+1}\phi^{(m\,2n)}_{0}+\frac{2 \e}{2n+3} \left(\frac{\hat{T}}{2}\right)^{2n+3}\phi^{(m\, 2n+1)}_{1} + \frac{2 \e^2}{2n+3} \left(\frac{\hat{T}}{2}\right)^{2n+3}\phi^{(m\, 2n)}_{2}\right]
\\
& + \e^{-1} \hat T \hat \zeta a^{\alpha \mu}a^{\beta \delta}Q_{\mu \delta} 
+\e \frac{\hat{T}^3}{12}( \hat{G} a^{\alpha\mu}a^{\beta\delta}-2 \hat{H} a^{\alpha\mu}\hb^{\beta\delta} - 4 \hat{H} \hb^{\alpha\mu}a^{\beta\delta} + \hb^{\alpha\mu}\hb^{\beta\delta} -2 a^{\alpha\mu}\hb^{\delta\gamma} \hb^\beta_\gamma) \, \hat \zeta Q_{\mu\delta}
\end{split}
\end{equation*}

In the limit $\e \to 0$, we only keep the terms corresponding to $m=2$ and $n=0$
\begin{equation*}
\begin{split}
\frac{n^{\alpha\beta}}{PL} = &  \,\hat{T} \phi^{(20)}_{0} + \frac{\hat{T}^3}{12} \left[ \phi_0^{(22)} +\phi_1^{(11)}+\phi_2^{(00)} \right] +\e^{-1}\hat{T} \hat \zeta a^{\alpha \mu}a^{\beta \delta}Q_{\mu \delta} + \e \frac{\hat{T}^3}{12} D^{\alpha\beta}
\\
 = & \,\hat{T}\left[2a^{\alpha\mu}a^{\beta\delta}\Delta^{(20)}_{\mu\delta}-a^{\alpha\beta} p^{(10)} \right]
 \\ 
 &+  \frac{\hat{T}^3}{12}\left[ -a^{\alpha\beta} p^{(12)}+2 a^{\alpha\mu}a^{\beta\delta} \Delta^{(22)}_{\mu\delta} -(\hb^{\alpha\beta}-2\hat{H}a^{\alpha\beta}) p^{(01)}
+4\, (a^{\beta\delta} \hb^{\alpha\mu} - \hat{H}a^{\alpha\mu}a^{\beta\delta})\Delta^{(11)}_{\mu\delta}  \right.
\\
& +\left.  
 (2 \hat{G} a^{\alpha\mu}a^{\beta\delta}+4 \hb^{\alpha\mu}\hb^{\beta\delta} - 4\hat H a^{\alpha\mu}\hb^{\beta\delta}-8a^{\alpha\mu}\hb^{\delta\gamma} \hb^\beta_\gamma -8 a^{\beta\delta}\hb^{\mu\gamma} \hb^\alpha_\gamma) \Delta^{(00)}_{\mu\delta}\right]
\\
& + \e^{-1} \hat{T} \hat\zeta  a^{\alpha \mu}a^{\beta \delta}Q_{\mu \delta}+\e \frac{\hat{T}^3}{12} D^{\alpha\beta},
\end{split}
\end{equation*}
where $\Delta^{(mn)}_{\alpha\beta}=\frac{1}{2}(u_{\alpha\vert\beta}^{(mn)}+u_{\beta\vert\alpha}^{(mn)})-\hb_{\alpha\beta} u_3^{(mn)}$, and $D^{\alpha\beta} = ( \hat{G} a^{\alpha\mu}a^{\beta\delta}-2 \hat{H} a^{\alpha\mu}\hb^{\beta\delta} - 4 \hat{H} \hb^{\alpha\mu}a^{\beta\delta} + \hb^{\alpha\mu}\hb^{\beta\delta} -2 a^{\alpha\mu}\hb^{\delta\gamma} \hb^\beta_\gamma) \hat \zeta Q_{\mu\delta}$.
\\

We now turn to the equation for the bending moment,
\begin{equation*}
\begin{split}
\frac{m^{\alpha\beta}}{PL^2}=& \, -\frac{1}{\e} \sum_{m,n=0}\varepsilon^m\int_{-\hat{T}/2}^{\hat{T}/2}\hat{z}^n\left[ \hz \phi^{(mn)}_{0}+\e\hz^2\phi^{(mn)}_{1}\right]d\hz -\e^3 \frac{\hat T^3}{12} (a^{\alpha\mu}\hb^{\beta\delta}+2a^{\beta\delta} \hb^{\alpha\mu}-2\hat{H}a^{\alpha\mu}a^{\beta\delta})\hat \zeta Q_{\mu\delta}
\\
= & -\,\frac{1}{\e} \sum_{m,n=0}\varepsilon^m \left[ \frac{2}{2n+3} \left(\frac{\hat{T}}{2}\right)^{2n+3}\phi^{(m \,2n+1)}_{0}+\frac{2 \e}{2n+3} \left(\frac{\hat{T}}{2}\right)^{2n+3}\phi^{(m\, 2n)}_{1}\right]
\\
&-\e \frac{\hat T^3}{12} (a^{\alpha\mu}\hb^{\beta\delta}+ 2a^{\beta\delta} \hb^{\alpha\mu}-2\hat{H}a^{\alpha\mu}a^{\beta\delta})\hat \zeta Q_{\mu\delta}
\end{split}
\end{equation*}

Again, keeping only the dominant term
\begin{equation*}
\begin{split}
\frac{m^{\alpha\beta}}{PL^2}& = -\frac{\hat{T}^3}{12}\left[ \phi^{(1 1)}_{0}+\phi^{(00)}_{1}\right]\\
&= -\frac{\hat{T}^3}{12}\left[  -a^{\alpha\beta} p^{(01)}+2 a^{\alpha\mu}a^{\beta\delta} \Delta^{(11)}_{\mu\delta} + 4\, (a^{\beta\delta} \hb^{\alpha\mu} - \hat{H}a^{\alpha\mu}a^{\beta\delta})\Delta^{(00)}_{\mu\delta} \right]
\\
&-\e \frac{\hat{T}^3}{12} (a^{\alpha\mu}\hb^{\beta\delta}+2a^{\beta\delta} \hb^{\alpha\mu}-2\hat{H}a^{\alpha\mu}a^{\beta\delta})\hat \zeta Q_{\mu\delta} 
\end{split}
\end{equation*}

The symmetric effective stress resultant $N^{\alpha\beta} = n^{\alpha\beta}-b^\beta_\lambda m^{\alpha\lambda}$ is 
\begin{equation}\label{eq:N_inc_}
\begin{split}
\frac{N^{\alpha\beta}}{PL}  = & \,\hat{T}\left[2a^{\alpha\mu}a^{\beta\delta}\Delta^{(20)}_{\mu\delta}-a^{\alpha\beta} p^{(10)} \right]
 \\ 
 &+  \frac{\hat{T}^3}{12}\left[ -a^{\alpha\beta} p^{(12)} +2 a^{\alpha\mu}a^{\beta\delta} \Delta^{(22)}_{\mu\delta} - 2(b^{\alpha\beta}-\hat{H}a^{\alpha\beta}) p^{(01)}
+ (4 a^{\beta\delta} \hb^{\alpha\mu} - 4 \hat{H}a^{\alpha\mu}a^{\beta\delta} + 2 a^{\alpha\mu} \hb^{\beta\delta} )\Delta^{(11)}_{\mu\delta}  \right.
\\
& +\left.  
 (2 \hat{G} a^{\alpha\mu}a^{\beta\delta}-8a^{\alpha\mu}\hb^{\delta\gamma} \hb^\beta_\gamma -8 a^{\beta\delta}\hb^{\mu\gamma} \hb^\alpha_\gamma) \Delta^{(00)}_{\mu\delta}\right]
\\
& + \e^{-1} \hat{T} \hat\zeta  a^{\alpha \mu}a^{\beta \delta}Q_{\mu \delta}+\e \frac{\hat{T}^3}{12} ( \hat{G} a^{\alpha\mu}a^{\beta\delta} - 4 \hat{H} \hb^{\alpha\mu}a^{\beta\delta} - \hb^{\alpha\mu}\hb^{\beta\delta} -3 a^{\alpha\mu}\hb^{\delta\gamma} \hb^\beta_\gamma) \hat \zeta Q_{\mu\delta}.
\end{split}
\end{equation}

The symmetric effective bending moment is given by,
\begin{equation}\label{eq:M_inc_}
\begin{split}
\frac{M^{\alpha\beta}}{L^2 P}& =  
-\frac{\hat{T}^3}{12}\left[  -a^{\alpha\beta} p^{(01)}+(a^{\alpha\mu}a^{\beta\delta}+a^{\beta\mu}a^{\alpha\delta}) \Delta^{(11)}_{\mu\delta} + 2\, [a^{\beta\delta} \hb^{\alpha\mu}+a^{\alpha\delta} \hb^{\beta\mu}  - \hat{H}(a^{\alpha\mu}a^{\beta\delta} + a^{\beta\mu}a^{\alpha\delta})]\Delta^{(00)}_{\mu\delta} \right]
\\
& - \e \frac{\hat{T}^3}{12}\left[\frac{1}{2}(a^{\alpha\mu}\hb^{\beta\delta}+a^{\beta\mu}\hb^{\alpha\delta})+a^{\beta\delta} \hb^{\alpha\mu}+a^{\alpha\delta} \hb^{\beta\mu}-\hat{H}(a^{\alpha\mu}a^{\beta\delta}+a^{\beta\mu}a^{\alpha\delta})\right]\hat \zeta Q_{\mu\delta}
\end{split}
\end{equation}

As for the membrane case in the main text, to obtain a close relation in terms of velocities referred to the midsurface, we need to find relations for the remaining coefficient. These relations are found using the expansion \eqref{eq:asymptoticbending} in the boundary conditions \eqref{eq:bc}, and the governing equations \eqref{eq:mass_conserv}, \eqref{eq:equilibrium}, and then equate to zero the factors in the expansion proportional to the same power in $\e$ and $\hat z$.

%
%

\paragraph{Boundary condition}
~~\\
The continuity of normal stress implies, from \eqref{eq:bc}, $\tau^{33}_\pm = P^3_{\pm}$. Then, with the expansions \eqref{eq:asymptoticbending},
\begin{equation*}
\begin{split}
\frac{1}{\e^3}\sum_{m,n=0} \varepsilon^m\hat{z}^n 
\left[-p^{(mn)} \e+  2 (n+1)u_3^{(m+1n+1)} \right] +\frac{1}{\e^2}\hat \zeta Q_{33} =  P^3_\pm.
\end{split}
\end{equation*}

Considering the term of order $\mathcal{O}(\e^{-3})$, $m = 0$, $n = 0$  we have that 
\begin{equation}\label{eq:u311_b_}
2u_3^{(11)} = 0.
\end{equation}

Next, from the term of order $\mathcal{O}(\e^{-2})$ we obtain for $n = 1$
\begin{equation}\label{eq:p01u322_}
p^{(01)} = 4 u_3 ^{22}.
\end{equation}

Finally, the $\mathcal{O}(\e^{-1})$ equation gives for $n = 0$
\begin{equation} \label{eq:p10u331_}
p^{(1 0)} = 2 u_3^{(31)},
\end{equation}
 
We now turn to the next boundary condition, relating the transverse load 
$\tau^{3\alpha}_\pm=2\mu g^{33}g^{\alpha \lambda}e_{3\lambda}=\vec{P}^{\alpha}_\pm$:
\begin{equation}
\begin{split}
\frac{P}{\e^3}\sum_{m,n=0} \varepsilon^m\hat{z}^n g^{\alpha\lambda}_+
\left[   (n+1)u_\lambda^{(m+1n+1)} +u_{3,\lambda}^{(mn)}+\hb^{\gamma}_\lambda u_\gamma^{(mn)}   -\e\hz  \,\hb^\gamma_\lambda u_\lambda^{(m+1n+1)} (n+1) \right] = \vec{P}^{\alpha}_\pm.
\end{split}
\end{equation}
The zero order term, $m=0, n=0$, in this equation gives,
\begin{equation}\label{eq:u11_bending_}
u_\lambda^{(11)} = - u_{3,\lambda}^{(00)} - \hb^{\gamma}_\lambda u_\gamma^{(00)}.
\end{equation}

\paragraph{Incompressibility: mass balance}

\begin{equation}
\begin{split}
\e^{-3}&\sum_{m,n=0}\varepsilon^m\hat{z}^n
 \left\lbrace   (n+1)u_3^{(m+1n+1)} +a^{\alpha\beta} u_{\alpha\vert\beta}^{(mn)}- 2\hat{H}u_3^{(mn)}   \right.
 \\
 +&\e\hz\left.\left[ -2\hat{H}(n+1) u_3^{(m+1n+1)}    +   2\hat{G}u_3^{(mn)} + (\hb^{\alpha\beta}-2\hat{H}a^{\alpha\beta})u_{\alpha\vert\beta}^{(mn)}\right] \right.
 +\left. \e^2\hz^2\hat{G}[u_3^{(m+1n+1)}]\right\rbrace 
 \\
 &=  \e^{-3}[-\hat k_d+\hat v_p\delta(\hat z-\hat T/2) 
 - \e \hat z \, 2 \hat H [-\hat k_d+\hat v_p \delta(\hat z-\hat T/2) ] + \e^2 \hat z^2 G [-\hat k_d+\hat v_p \delta(\hat z-\hat T/2) ].
\end{split}
\end{equation}

Consider first the case $(m,n)=(0,0)$, giving $\, u_3^{(11)} +  a^{\alpha\beta}\Delta_{\alpha\beta}^{(00)} + \hat k_d- \frac{\hat v_p}{\hat T}=0 $.  Since $u_3^{(11)} = 0$ it means that $a^{\alpha\beta}\Delta_{\alpha\beta}^{(00)} =\frac{\hat v_p}{\hat T} - \hat k_d$. 

Next, consider the pair $(m, n) = (2, 0)$, leading to $u_3^{(31)} =-a^{\alpha\beta}\Delta_{\alpha\beta}^{(20)}$.
Then, using Eq.~\eqref{eq:p10u331_}, we obtain
\begin{equation}\label{eq:p10_}
p^{(10)}= - 2a^{\alpha\beta}\Delta_{\alpha\beta}^{(20)}.
\end{equation}

If we consider $m=1$, and $n=1$, we can show that $2u_3^{(22)} = a^{\alpha\beta}\Omega^{(00)}_{\alpha\beta} + 2\hat{H} \left(\hat k_d - \frac{\hat{v}_p}{\hat T} \right)$.
It follows from Eq.~\eqref{eq:p01u322_} that
\begin{equation}\label{eq:p01_}
p^{(01)} = 2a^{\alpha\beta}\Omega^{(00)}_{\alpha\beta}+ 4\hat H\left( \hat k_d - \frac{\hat v_p}{\hat T}\right).
\end{equation}
Finally, the relation for $m=2$, and $n = 2$ leads to $3 u_{3}^{(33)} = (\hat{H} a^{\alpha\beta} + \hb^{\alpha\beta})\Omega_{\alpha\beta}^{(00)} + 3 (2 \hat{H}^2-\hat{G} )\left( \hat k_d - \frac{\hat v_p}{\hat T}\right)$.

\paragraph{Equilibrium equations}
~~\\
The remaining parameter $p^{(12)}$ can be found from the out-of-plane equilibrium equation \eqref{eq:equilibrium_out_of_plane} when looking at the $\mathcal{O}( \e^{-2}, \hat z)$ term
\begin{equation}\label{eq:p12_}
p^{(12)}=\hb^{\alpha\beta}\Omega_{ \alpha\beta}^{(00)}+4( \hat{H}^2-\hat{G} )\left( \hat  k_d - \frac{\hat v_p}{\hat T}\right)
\end{equation}

We now substitute Eq.~\eqref{eq:u11_bending_}, \eqref{eq:p10_}, \eqref{eq:p01_}, \eqref{eq:p12_} into Eq.~\eqref{eq:N_inc_}, and \eqref{eq:M_inc_} to obtain the dimensional (symmetric) effective stress and bending moment
\begin{equation}
\begin{split}
N^{\alpha\beta} & =\, 4 \mu T \mathcal{A}^{\alpha\beta\mu\delta}\Delta_{\mu\delta} + T \zeta  a^{\alpha \mu} a^{\beta \delta} Q_{\mu \delta}
\\
& + \mu \frac{T^3}{12}\left[ C^{\alpha\beta\mu\delta}\Omega_{\mu\delta} + B^{\alpha\beta} \left( k_d - \frac{ v_p}{ T}\right)\right]
\\
& + \frac{T^3}{12}( G a^{\alpha\mu}a^{\beta\delta} - 4 H b^{\alpha\mu}a^{\beta\delta} - b^{\alpha\mu} b^{\beta\delta} -3 a^{\alpha\mu} b^{\delta\gamma}  b^\beta_\gamma)  \zeta Q_{\mu\delta},
\end{split}
\end{equation}
    where   $\mathcal{C}^{\alpha\beta\mu\delta}=8H\mathcal{A}^{\alpha\beta\mu\delta}-a^{\alpha\beta}b^{\mu\delta}-5b^{\alpha\beta}a^{\mu\delta}-\frac{3}{2}(a^{\alpha\mu}b^{\beta\delta}+a^{\alpha\delta} b^{\beta\mu}+a^{\beta\mu}b^{\alpha\delta}+a^{\beta\delta} b^{\alpha\mu})$, and $B^{\alpha\beta} = 2 (G -H^2) a^{\alpha\beta} +6 H b^{\alpha\beta} +10 b^\alpha_\mu b^{\beta\mu}$.
\\

The effective bending moment,
\begin{equation}
\begin{split}
M^{\alpha\beta} =
\mu \frac{T^3}{3}\mathcal{A}^{\alpha\beta\mu\delta}\Omega_{\mu\delta} + \mu \frac{T^3}{12}(6H a^{\alpha\beta} + 4b^{\alpha\beta})\left(k_d - \frac{v_p}{T}\right)
+\frac{T^3}{12}\mathcal{D}^{\alpha\beta\mu\delta} Q_{\mu\delta}\zeta
\end{split}
\end{equation}
Where  $\mathcal{D}^{\alpha\beta\mu\delta}=H(a^{\alpha\mu}a^{\beta\delta}+a^{\beta\mu}a^{\alpha\delta})+a^{\alpha\beta}\hb^{\mu\delta}-\frac{3}{4}(a^{\alpha\mu}\hb^{\beta\delta}+a^{\alpha\delta}\hb^{\beta\mu}+a^{\beta\mu}\hb^{\alpha\delta}+a^{\beta\delta}\hb^{\alpha\mu})$.
\FloatBarrier

\section{Numerical values}

The physical parameters used in the main text are summarized in Table \ref{table}.

\begin{table}[H]
\begin{tabular}{|l|l|l|l|l|}
\hline
Notation & Quantity & \begin{tabular}[c]{@{}l@{}}Experimental\\ Value\end{tabular} & \begin{tabular}[c]{@{}l@{}}Numerical \\ Value\end{tabular} &  Ref(s). \\\hline
$R$               & Cell radius            & 10 -100$\mu$m         & 1   & \cite{bement2005rhoA,von2009action} \\ \hline
$T$               & Cortex thickness       & 0.2 - 2$\mu$m         & 0.02 & \cite{clark2013monitoring, hiramoto1957thickness}\\ \hline
$w$               & Rho-A-GTP signal width & 1 - 10$\mu$m          & 0.15 & \cite{bement2005rhoA,von2009action} \\ \hline
$\zeta$           & Basal Active stress    & $10^3$Pa              & 1    & \cite{tinevez2009role}\\ \hline
$t_a$             & Cytokinesis timescale  & $10^2$s               & 1   & \cite{bement2005rhoA,von2009action} \\ \hline
$\mu = \zeta t_a$ & Actomyosin shear viscosity   & $10^5$Pa s           & 1  & \\  \hline
$k_d$             & Depolymerization       & 0.04 $s^-1$           & 4  & \cite{guha2005cortical} \\ \hline
$v_p = T k_d$             & Polymerization         & 0.008 - 0.08 $\mu$m/s & 0.08 &  \\ \hline
\end{tabular}
\caption{Numerical values have been chosen so that $R_0 = 1$ and $T_a = 10$ are the reference length and timescale.}\label{table}
\end{table}
\FloatBarrier

\section{Boundary conditions}\label{app:bc}
    To apply the boundary conditions we need to define a set of basis vector tangent and normal to the outer surface. The position vector in the outer surfaces is \[\vec{x}^{\pm}=\bm{x}(\xi^\alpha,t)\pm \frac{1}{2}T(\xi^\alpha,t)\vec{n}(\xi^{\alpha},t). \] Here we specified that the thickness $T$ is a function of time and varies along the sheet. Moreover, $\pm$ indicates that the variable in question is evaluated at $z=\pm T/2$. Then, the covariant and contravariant tangent vectors to the outer surfaces are 
\begin{equation}\label{eq:outer_surface_basis}
    \vec{c}^{\pm}_\alpha\equiv\pm \vec{x}^\pm_{,\alpha}=\pm \g^\pm_\alpha+\frac{1}{2}T_{,\alpha}\g_3, \quad \vec{c}^\alpha_\pm=\pm\g^\alpha_\pm+\frac{g^{\alpha\lambda} T_{,\lambda}}{2\Lambda^2_\pm}(\g^3\mp\frac{1}{2}T_{,\beta}\g^\beta_\pm),
\end{equation}
where $\Lambda_\pm=(1+\frac{1}{4}g^{\alpha\beta}_\pm T_{,\alpha}T_{,\beta})^{1/2}$. The outward unit normal vectors are $\vec{c}^\pm_3=\vec{c}^3_\pm\equiv \vec{n}^\pm = n_i^\pm\g^i_\pm$, with $n^\pm_\alpha=-\Lambda^{-1}_\pm \frac{1}{2}T_{,\alpha} $ and $n^\pm_3=\pm\Lambda^{-1}_\pm$. 
    
    The main driver of cortex deformation is generally the local motor activity, which creates additional internal contractile stress (akin to a negative pressure) in the layer, that cells can control longitudinally. However, external normal forces are also present in the form of hydrostatic pressure stemming from the extracellular or intracellular medium (or cytoplasm). Tangential external loads may also exist at the inner and outer surfaces of the cortex, such as friction forces arising from the interaction with the plasma membrane or viscous resistance from the cytoplasm. Then, let $\vec{P}^{\pm} = P^{\pm}_i \vec{c}^i_{\pm} = P_{\pm}^i \vec{c}_i^{\pm}$ represent the stress vector applied to the outer surfaces of the cell, per unit area of those outer surfaces. The vectors $\vec{c}^{i}_\pm$ and $\vec{c}_{i}^\pm$ are the given by \eqref{eq:outer_surface_basis}. The tangential components of the applied stress are $P^{\pm}_\alpha$ or  $P_{\pm}^\alpha$, and the the normal components $P^\pm_3$ or $P_\pm^3$. Continuity of the stress vector at the outer surfaces require
    \begin{equation}\label{eq:bc}
        \tau^{ij}_\pm n_i^\pm \vec{g}^\pm_ j=\vec{P}^{\pm}.
    \end{equation}
    Notice, however, that the boundary stresses appear in the balance equations only through terms in $F^j_{\pm} = \Lambda_{\pm}\sigma^{ij}_{\pm}n^{\pm}_i$. To obtain an expression for $F^j_{\pm}$ in terms of $\vec{P}^{\pm}$ we project \eqref{eq:bc} onto the midsurface base vectors,
    \begin{subequations}
    \begin{align}
        &F^3_\pm = \pm h_{\pm} P^3_\pm + \frac{1}{2}T_{,\alpha} P^\alpha_\pm +\mathcal{O}(\e^2 |\vec{P}|)
        \\
        &F^\alpha_\pm = \pm h_{\pm} P^\alpha_\pm - \frac{1}{2}T b^{\alpha}_{\beta} P^{\beta}_{\pm} - \frac{1}{2} a^{\alpha\beta}T_{,\beta} P^3_\pm  +\mathcal{O}(\e^2 |\vec{P}|).
    \end{align}
    \end{subequations}



\section{On the thin-shell limit of the shearable model}
In this section, we show how to retrieve Eq.~\eqref{eq:Delta}, and \eqref{eq:omega}, from their shearable counterparts Eq.~\eqref{eq:discr-Delta_ind}, and \eqref{eq:discr-Omega_ind} in the thin shell limit where shear is negligible.

Lets first decompose the midsurface velocity as $\bm{U} = U^\lambda \bm{a}_\lambda + U^3 \bm n$. Then, using the Gauss-Weingarten relations
we can show that
\begin{equation}\label{eq:U_alpha}
    (U^\lambda \bm{a}_\lambda + U^3 \bm n)_{,\alpha} = (U_{\lambda|\alpha} - b_{\lambda\alpha}U_{3})\bm{a}^\lambda + (U_{3,\alpha} + b^\lambda_{\alpha}U_{\lambda})\bm{n}.
\end{equation}

Using this expression into Eq.~\eqref{eq:discr-Delta_ind} and recalling that $\bm{X}_{,\alpha} = \bm{a}_\alpha$ will lead directly to Eq.~\eqref{eq:Delta}.

Next, note that when $\gamma_\alpha = 0$, we have $\bm{\omega} \bm{X}_{,\alpha} = - \bm{d}^0 \bm{U}_{,\alpha}$. Since $\bm{d}^0 = \bm{n}$, we can directly write the expression for $\bm{\omega}$ as 
\begin{equation}\label{eq:ddot}
    \bm{\omega} =  - (U_{3,\alpha} + b^\lambda_{\alpha}U_{\lambda})\bm{a}^{\alpha}.
\end{equation}
The direct substitution of the above equation into Eq.~\eqref{eq:discr-Omega_ind}, the use of the Gauss-Weingarten and Codazzi ($b^\lambda_{\alpha| \beta}= b^\lambda_{\beta |\alpha}$) relations, will lead to Eq.~\eqref{eq:omega}.

\end{document}